\documentclass[useAMS,usenatbib]{mn2e}
\usepackage{natbib, graphicx, times}
 \usepackage{graphicx}
 \usepackage{subfigure}
\usepackage{stackrel}
\usepackage{color,hyperref}

\usepackage{amssymb,amsmath}
\usepackage{bm}
\def\red#1 {\textcolor{red}{#1}\ }   
\def\blue#1 {\textcolor{blue}{#1}\ }   

\newcommand{\apj}{ApJ}
\newcommand{\apjl}{ApJ}
\newcommand{\apjs}{ApJS}
\newcommand{\aap}{A \& A}
\newcommand{\araa}{ARA\&A}
\newcommand{\aj}{AJ}
\newcommand{\mnras}{MNRAS}

\newcommand{\nat}{Nature}

\newcommand{\icarus}{Icarus}

\title[Orbit evolution in disc encounters]{Stellar orbit evolution in close circumstellar disc encounters}

\author[Mu\~noz et al.]{D. J. Mu\~noz$^{1,2}$\thanks{E-mail: dmunoz@astro.cornell.edu},
 K. Kratter$^{3}$,
 M. Vogelsberger$^{1}$, 
 L. Hernquist$^{1}$
and~V. Springel$^{4,5}$\\
$^{1}$ Harvard-Smithsonian Center for Astrophysics, 60 Garden Street, Cambridge, MA 02138 ,USA\\
$^{2}$ Center for Space Research, Department of Astronomy, Cornell University, Ithaca, NY 14853, USA\\
$^{3}$ Steward Observatory, University of Arizona, 933 North Cherry Ave, Tucson, AZ, 85721, USA \\
$^{4}$ Heidelberg Institute for Theoretical Studies, Schloss-Wolfsbrunnenweg 35, 69118 Heidelberg, Germany \\
$^{5}$ Zentrum f\"{u}r Astronomie der Universit\"{a}t Heidelberg, ARI, M\"onchhofstr. 12-14, 69120 Heidelberg, Germany}

\begin{document}

\pagerange{\pageref{firstpage}--\pageref{lastpage}} \pubyear{2014}

\maketitle

\label{firstpage}

\begin{abstract}
The formation and early evolution of circumstellar discs often occurs within dense, newborn stellar clusters.
For the first time, we apply the moving-mesh code {\footnotesize AREPO}, to circumstellar discs in 3-D, focusing
on disc-disc interactions that result from stellar fly-bys. Although a small fraction
of stars are expected to undergo close approaches, the
outcomes of the most violent encounters might leave an imprint on the discs and host stars that
 will influence both their orbits and their ability to form planets. We first construct well-behaved
  3-D models of self-gravitating discs, and then create a suite of numerical 
experiments of parabolic encounters, exploring the effects of  pericenter separation $r_p$, disc 
orientation and disc-star mass ratio ($M_d/M_*$) on the orbital evolution of the host stars. Close 
encounters ($2r_p\lesssim$ disc radius) can truncate discs on very short time scales.
If discs are massive, close encounters facilitate enough orbital angular momentum 
extraction to induce  stellar capture. 
We find that for realistic primordial disc masses $M_d\lesssim0.1M_*$,
non-colliding encounters induce minor orbital changes,  which is  consistent 
with analytic calculations of encounters in the linear regime. The same disc masses produce entirely
different results for grazing/colliding encounters. In the latter case, rapidly cooling discs 
lose orbital energy by radiating away the energy excess of the shock-heated gas, thus causing
capture of the host stars into a bound orbit. In rare cases, a tight binary with a circumbinary disc forms as a result of
this encounter.
\end{abstract}

\begin{keywords}
hydrodynamics -- methods: numerical -- planets and satellites: formation -- protoplanetary discs -- binaries: general.
\vspace{-0.25in}
\end{keywords}


\section{Introduction}

The variety of processes that influence circumstellar disc evolution 
in multi-star systems are not yet well understood.  
discs can be influenced by the primordial cluster kinematics, with
important consequences for planet formation.  In contrast
with observations of  more evolved Class II objects (for which there is evidence that multiplicity 
can have a strong influence on disc sizes and morphologies) Class 0 objects 
are more elusive from direct observation \citep[e.g.][]{jor09,che13,chi12,tob13,mur13}. Nevertheless, the  existence of
evolved discs in binary systems implies that earlier, massive rotating structures
must have interacted in some way with stars other than their host.
Exploring these more violent interactions, where pericenter separation becomes comparable to disc size,
requires high resolution, three dimensional numerical simulations.

The role of stellar flybys in planet formation has been explored from the point
of view of planet stability \citep[e.g.][]{ada01,ada06,duk12} through exhaustive Monte Carlo $N$-body
methods. The detailed effects of stellar flybys on single gas discs have also been explored 
through isolated simulations  \citep[e.g.][]{for09,she06,moe06,she10,thi10}. Close passages
can truncate discs, which affects planet formation, and also trigger gravitational instabilities in discs
\citep[e.g][]{she10,thi10}.  Nearby stars also accelerate disc photoevaporation 
even when their dynamical influence is small \citep{and13}.

Bound multiple systems affect
 disc evolution and planet formation  radically. 
The perturbations from a stellar-mass companion lead to tidal truncation
\citep[e.g.,][]{art94}, warping/bending \citep[e.g.,][]{lar96,ogi01} or 
hastened disc dispersal \citep[e.g.,][]{ale12,krau12a,har12}.
Recent observations  of pre-main sequence stars in binary systems
 have directly confirmed the importance of some of these effects.
disc size has been shown to depend strongly on stellar separation \citep{har12} and close
passages have been associated with major disruptions in disc geometry, although these might require
additional physics (e.g. winds, external photo-evaporation) other than
the gravitational interactions between the disc and a companion star \citep[see e.g.,][]{cab06,sal14}.

In principle, the higher order multipole moments of a massive circumstellar disc's
potential could modify stellar encounters. Although disc masses
of pre-main sequence stars (e.g. Class II young stellar objects) are not expected to exceed $1\%$ of the 
host star's mass \citep[e.g.][]{and07a,and09}, very early protostellar objects are expected to have more massive 
discs or envelopes. The role of these massive young discs/envelopes on stellar dynamics has not been
explored much theoretically, but it is possible that the gas component plays
a short lived, but important role that cannot be captured by simple collisionless dynamics.

Despite the increased sophistication of cluster
models \citep{ada06,par12,bat11,bat12,cra13},
the dynamical range required on timescales of $N$-body systems of
stars with respect to their planetary systems makes direct simulation of
full clusters prohibitive computationally. Instead, we isolate two-body systems in order to study disc-disc interactions
at high resolution.

In this work, we focus on the direct simulation of circumstellar disc flybys, solving the
equations of three-dimensional, self-gravitating hydrodynamics discretized over a moving
Voronoi mesh as implemented by the {\footnotesize AREPO} code \citep{spr10a}.
In Section~\ref{sec:experiments}, we describe the numerical set-up, detailing the individual
disc models and the orbital configuration of the disc encounters. We summarize the results
of the numerical experiments in Section~\ref{sec:results}, focusing on the evolution of
the stellar orbits. In Section~\ref{sec:discussion}, we discuss the consequences of these disc
encounters and the plausibility and rate of their occurrence. 

\section{Numerical experiments on disc-disc interaction}\label{sec:experiments}
\subsection{Previous work}

Numerical experiments with isolated configurations of star-disc and disc-disc interaction 
on spatial scales of $\sim100$~AU enable the detailed study of regions that are usually unresolved 
in self-consistent, {\it ab initio} simulations of star forming clouds with spatial scales of $\sim10^5$~AU.
Note that even the state-of-the-art star formation simulations of \citet{bat12} -- which form discs
around protostars -- have an SPH particle mass of $1.43\times10^{-5}M_\odot$, implying that a $0.01M_\odot$ disc
is composed of barely 700 resolution elements. 

Despite the rareness of events like direct star-disc and disc-disc interactions, such
encounters have received significant attention in the literature, either focusing on their role in
the tidal evolution of a binary and orbital capture \citep{cla91a,cla91b,ost94}, or
by studying the possible triggering of spiral arms, gravitational instability (GI) and fragmentation
\citep{bof98,lin98,pfa03,pfa05,pfa07,for09,she10}.

Studies of disc-disc collisions by direct numerical simulation 
date back to \citet{lin98} and \citet{wat98a,wat98b} \citep[see also][]{bof98}, where authors hypothesized that
condensation of material in tidally induced tails could produce a population of brown dwarfs.
This idea has been revisited both with pure $N$-body approaches \citep{pfa03,pfa05,thi05,pfa07}
and with gas dynamics \citep{for09,she06,she10,thi10}. All the hydrodynamical
studies included self-gravity, although some of the $N$-body ones used test particles \citep[e.g.][]{pfa05},
only focusing on the passive response of the disc to an external perturber. 
These studies have concentrated on the tidal generation of spiral arms \citep[e.g.][]{pfa03}; on disc fragmentation
and formation of substellar mass objects \citep[e.g.][]{lin98,she06,she10,thi10} and disc
truncation \citep[e.g.][]{for09} and, more recently, on the stability of disc-embedded planets \citep{pic14}.
Those studies that include gas  were all carried out using smoothed particle hydrodynamics
\citep[SPH;][]{luc77,gin77,mon92,spr10b}. 

The fact that SPH has been the method of choice for disc-disc interaction
reflects the necessity of having a geometrically flexible  numerical scheme
(e.g., without preferred orientations/symmetries).
Isolated discs are nearly axisymmetric, and thus benefit
from the use of structured grids in cylindrical coordinates, since such grid
configurations favor low numerical diffusivity for azimuthal flow. As soon as this symmetry is broken (e.g., by combining two circumstellar discs
moving at supersonic speeds toward each other), the benefit of structured grids becomes less clear.
Consequently, this problem is ideal for a quasi-Lagrangian 
Eulerian code like {\footnotesize AREPO} \citep{spr10a}, in which the control volumes evolve and move
in a similar way to SPH particles; i.e., by following the local velocity field. Yet by being locally a grid
code,  {\footnotesize AREPO} does not suffer from some of the numerical artifacts SPH is known
to develop, such as clumping instabilities, suppression of hydrodynamic instabilities, artificial surface tension,
zeroth order error terms, etc \citep[see][]{vog12,sij12,bau12,deh12}. One of the main advantages of using a code like {\footnotesize AREPO}
for circumstellar disc simulations is the minimization of the so-called 
``high Mach number problem". When solving the Euler equation
in the rest frame of a moving cell, the local velocity of the flow is irrelevant
for the outcome of the calculation. This can be a great advantage when
simulating moving discs in binary or multiple stellar systems. 

The study closest to our work is \citet{she10}, where the authors performed fully self-gravitating
simulations of gas discs encounters using SPH. Similarly to \citet{lin98}, \citet{she10} focused on the 
formation of self-gravitating objects (in the brown dwarf range) in tidal tails \citep[reminiscent of the work of][]{bar92}
 and in compression shocks during encounters of {\it very} massive discs (the values of star mass, disc
  mass and disc radius  used are $M_*=0.6M_\odot$, $M_d=0.4M_\odot$
and $R_d=250$~AU in \citealp{lin98}, and $M_*=0.5M_\odot$, $M_d=0.6M_\odot$ and $R_d\sim1000$~AU in \citealp{she10}).
In this work, we model discs of more moderate -- and more plausible -- masses ($10\%$ of the mass of the star),
focusing on the role of the tidal forces on the orbital evolution of the host stars, and exploring how small the impact parameter
must be in order to cause a significant change to the original orbits. 

\begin{figure*}
\begin{center}
\includegraphics[width=1.02\textwidth]{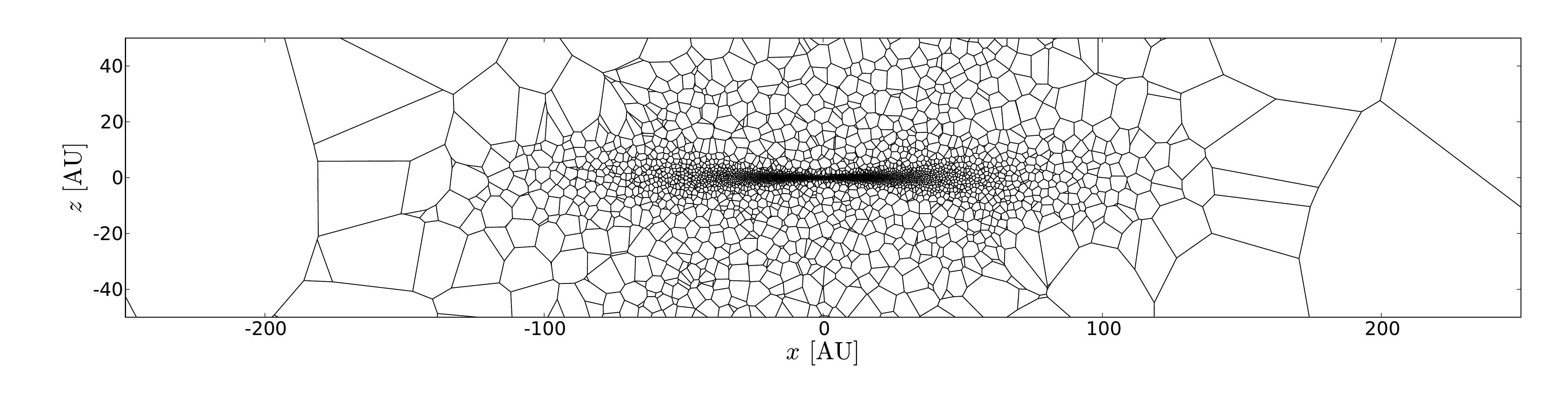}
\vspace{-0.4in}
\caption{Slice through the mesh of a disc model, showing the transition
between a mass-based sampling within  the disc and a volume-based sampling in the background.
\label{fig:mesh_slice}
\vspace{-0.1in}}
\end{center}
\end{figure*}

\subsection{The {\small{AREPO}} code}
{
{\footnotesize AREPO} \citep{spr10a,spr11} is a second-order finite-volume Godunov code
that solves the equations of hydrodynamics on a moving mesh. This approach is
built upon the idea that control volumes can be identified with the polygons/polyhedra of
a Voronoi tessellation, which in turn is constructed from a set
of  {\it moving} mesh-generating points. Since a Voronoi tessellation changes
continuously for smooth spatial trajectories of the generating points, each
cell can be regarded as a ``quasi-Lagrangian fluid parcel" \citep[e.g., see][]{vog12}.
Besides ideal hydrodynamics coupled to self-gravity, the code currently includes 
routines for magneto-hydrodynamics (\citealp{pak11,pak13}; see also \citealp{moc14a}
and \citealp{moc14b}),
physical viscosity \citep{mun13}, radiative transfer in different approximations 
\citep{pet11,sale14}, as well as numerous sub-resolution prescriptions for galaxy formation
physics \citep{vog13}.}

{As in any unsplit, second-order, finite-volume method \citep[see e.g.,][]{tor09},
the hyperbolic Euler equations are discretized for the $i$-th cell as \citep{spr10a}:}
\begin{equation}\label{eq:finite_volume}
\mathbf{Q}_i^{(n+1)}= \mathbf{Q}_i^{(n)}-\Delta t \sum_j A_{ij} \hat{\mathbf{F}}_{ij}^{(n+1/2)}~~,
\end{equation}
{where $\mathbf{Q}$ is the volume integral of the state vector $\mathbf{U}=(\rho,\rho\mathbf{v},\rho e)$
(i.e., mass, momentum and energy densities), $A_{ij}$ is the area of the face shared by the 
$i$-th and $j$-th cells and $\hat{\mathbf{F}}_{ij}^{(n+1/2)}$ is a time-centered, area-averaged estimate of the
flux vector $\mathbf{F}_{ij}$, which is an analytic function of the fluid variables.}

{However, since the Voronoi cells are moving, the flux $\mathbf{F}_{ij}$ exchanged 
between the $i$-th and $j$-th cells needs to take into account the
velocity $\mathbf{w}$ of a moving boundary: }
\begin{equation}
\begin{split}
\mathbf{F}_{ij}&=\frac{1}{A_{ij}}\int_{A_{ij}}\left[\mathbf{F}(\mathbf{U})-\mathbf{U}\mathbf{w}^T\right]d\mathbf{A}_{ij}\\
&\approx \hat{\mathbf{F}}_{ij}= \mathbf{F}(\mathbf{U}_{ij}) -\mathbf{U}_{ij}\mathbf{w}_{ij}^T
\end{split}
\end{equation}
{where the state vector estimate $\mathbf{U}_{ij}$ is obtained from the solution
of a Riemann problem at the centroid of the interface and $\mathbf{w}_{ij}$
is the velocity of the face centroid. This Riemann
problem is solved on the frame of the moving face, and thus errors
associated with this operation are independent of the magnitude of 
$\mathbf{w}_{ij}$. As a consequence, the truncation error in
Equation~(\ref{eq:finite_volume}) is velocity independent, a property
that guarantees that the diffusivity and numerical noise in our
disc simulation is independent of how fast they are moving across
the computational domain. In addition, this approach allows for longer
time steps, since the Courant condition is applied in the frame of the
moving cell, and not in lab frame coordinates. This minimizes the
numerical diffusion that would arise from taking many time-steps per
dynamical time.}

{Gravity in {\footnotesize AREPO} is included as a source term to the right hand side
of Equation~(\ref{eq:finite_volume}). Integration of this term is carried out by a modified
operator-splitting approach \citep{spr10a}.
For self-gravitating systems, the gravitational force is calculated using
a hierarchical octree algorithm \citep{bar86}. All the simulations presented in this work
are self-gravitating.}

{Because of the discretization of the equations of motion in cartesian coordinates, and
the use of an octree algorithim for the gravitational force,} in its
current formulation {\footnotesize AREPO} does not conserve angular momentum
to machine precision, but future modifications
to the {\footnotesize AREPO} algorithm may improve the code's performance
in this regard, enabling integration for longer timescales than the ones used
in this work
\footnote{See Pakmor, Mu\~noz and Springel (in preparation) for a detailed analysis of the conservation and diffusion 
of angular momentum in moving-mesh hydrodynamics, including new schemes
that may improve the code's performance in this regard.
}.  {For the disc models used here 
(see Section~\ref{sec:disc_models} below) and integration times of our
runs, angular momentum conservation is satisfied to within $1\%$, which should not strongly influence our conclusions.  Note that a significant amount of the observed angular momentum variability
is due to the accretion of gas by the stars (we use a rather aggressive sink particle
algorithm acting over a sphere of radius $r_\mathrm{acc}=1$~AU around each
star), which amounts to $\dot{M}_*\gtrsim10^{-7}M_\odot$~yr$^{-1}$.}
A benefit of the moving-mesh approach is that, this error in angular momentum
is independent of the disc bulk motion, which allows
us to experiment with encounters at different speed without the need of adapting
the resolution accordingly.

\subsection{Circumstellar disc models}\label{sec:disc_models}
The moving-mesh code {\footnotesize AREPO} was benchmarked by \citet{mun14a} for circumstellar disc simulations
in 2-D, finding good agreement with other popular grid codes used for two-dimensional calculations of planet-disc 
interaction. In this work, we tackle the much more challenging problem of modeling full three-dimensional
discs of finite extent in stationary equilibrium.
We use a basic, observationally-justified model, which consists of a self-consistent, self-gravitating solution
for a disc with a surface density distribution that satisfies the Lynden-Bell--Pringle model \citep{lyn74}:
\begin{equation}\label{eq:surface_density}
\Sigma(R)=(2-p)\frac{M_d}{2\pi R_c^2}\left(\frac{R}{R_c}\right)^{-p}\exp\left[-\left(\frac{R}{R_c}\right)^{2-p}\right]~~,
\end{equation}
where $M_d=0.05\,M_\odot$, $p=1$, $R_c=20$~AU and a stellar mass of  $M_*=0.45$ is used for all our 
numerical experiments unless otherwise noted. {These characteristic radii are consistent with
the protoplanetary disc surface density profiles fitted from submillimeter observations
by \citet{and09} using Equations~(\ref{eq:surface_density})
(these authors find a wide spread in $R_c$, going from $\sim$20 to $\sim$200~AU).
Our models are comparable in size and
mass to those by \citet{for09}, who chose outer disc radii of 40~AU and star and disc masses of $M_*=0.5M_\odot$
and $M_d=0.07M_\odot$ respectively.}

{Although the disc's characteristic radius $R_c$ 
(Equation~\ref{eq:surface_density} defines the scale at which the disc surface density profile transitions from 
power-law form to an exponential cutoff), it does
not define a specific disc {\it size}. We define $R_d$ simply as the radius that encloses $95\%$ of the disc mass. For a surface
density profile given by Equation~(\ref{eq:surface_density}), it can be shown that the enclosed 
mass at radius $R$ is}
\begin{subequations}
\begin{align}
\label{eq:enclosed_mass}
M_d(<R)&=M_d\left\{1-\exp\left[-\left(\frac{R}{R_c}\right)^{2-p}\right]\right\}~~,\\
\label{eq:disc_radius}
&\text{  with   }M_d(<R_d)\equiv 0.95 M_d~~.
\end{align}
\end{subequations}
{For $p=1$, $R_d\approx3R_c$, i.e. the disc size is approximately 60~AU in our models.}

Simulations are carried out by discretizing the 3-D volume
into cells (Voronoi polyhedra) with a mass resolution of $9\times10^{-7}\,M_\odot$ (0.3 Earth masses)
except for those dilute regions outside the discs, where a cell volume ceiling is imposed. This implies
that our fiducial $0.05\,M_\odot$ discs 
are described by $\sim500000$ cells, while more massive discs contain a larger
number of resolution elements. Figure~\ref{fig:mesh_slice} shows a vertical slice of the Voronoi mesh, crossing the disc
at its centre. The smooth transition from a dense mesh into a coarse background mesh can be clearly seen.
A detailed explanation of the initial conditions and mesh-generation can be found in Appendix~\ref{app:models}.
{These initial conditions are very quiet: when discs are evolved in isolation, the projected surface density remains to be well
represented by Equation~\ref{eq:surface_density} over a significant fraction of the integration time of our
simulations (5000 years). Therefore, a relaxation stage of our initial conditions (which is very common in SPH simulations)
is not necessary and it is not carried out here.}

In addition to the density profile of Equation~(\ref{eq:surface_density}), a temperature profile is required
to completely specify the vertical structure of the disc. For all models presented in this work,
the temperature profile follows a power law $T\propto R^{-l}$ with fixed $l=1/2$, which corresponds 
to a mildly flared disc. The aspect ratio of the disc as a function of radius is set by the normalization of
the temperature profile. {The sound speed profile is a {\it fixed} function of
stellocentric distance, and is satisfied at all times as the discs move
across the computational domain \citep[as done by][]{she10}. We have employed two ways of
enforcing this ``local isothermality". One way is to ignore the energy conservation equation
and solve the Euler equations (\ref{eq:finite_volume}) using an iterative isothermal
Riemann solver with a {\it space-varying} sound speed \citep[e.g., as done in][]{mun14a}.
An alternative is to solve the energy equation with an ideal-gas equation of state
setting the adiabatic index to be $\gamma=1+\epsilon$ with $\epsilon\ll1$. Of course, this
is a reasonable shortcut if the discs do not exchange mass during the encounter, since this
approximation implies that gas parcels preserve their initial temperature along Lagrangian
trajectories: gas is not allowed to heat (or cool) if its stellocentric distance changes
\citep[see discussion on this issue in][]{mun14a}. In addition, $\epsilon$ needs to be small
enough to counteract adiabatic shock heating. For example, for gas shocked at a Mach number
of 30, $\epsilon$ needs to be smaller than $3\times10^{-5}$ for the post-shock temperature
to be within $1\%$ of the pre-shock temperature
\footnote{For a gas with adiabatic index $\gamma=1+\epsilon$ and upstream temperature $T_1$,
the temperature jump condition given a Mach number $M_1$ is
\begin{equation}
T_2/T_1\approx \frac{1}{2}\left(\epsilon M_1^2+2\right)~~.
\end{equation}
}.
We have tested our runs using $\gamma=1+10^{-5}$, plus a temperature relaxation term 
of the form $\dot{T}=-(T-T_0)/\tau_\mathrm{cool}$ 
with a cooling time $\tau_\mathrm{cool}$ equal to the local time-step. This additional source term in the energy
equation allows for quick cooling/heating toward a reference temperature $T_0$ that is set
by the distance to the closest star. The benefit of this second approach is that more
general/realistic cooling functions can be tested by simply changing $\gamma$ and the cooling time,
such that cooling/heating is delayed rather than ``instantaneous". For the results presented in this work, 
no significant difference has been measured when trying these two different cooling approaches.
}

Our numerical scheme calculates the gas self-gravity and thus is able to capture disc fragmentation
and object formation. However, the disc masses involved in our study make fragmentation much less
likely than for the models of \citet{she10}. 
None of the simulations performed for this work produce long-lived fragments. 
We have checked that, with cooler discs (temperatures scaled down by a factor or two or more), 
fragmentation is indeed possible (note that isothermal gas can be compressed by shocks
to arbitrarily high densities).
However, our choice of 
temperature is physically motivated and consistent with observations  \citep[see, e.g.,][]{kra08,kra10b}.
The disc scale height is reasonably small ($H/R=0.1$
at the disc characteristic radius) and the temperatures only reach a maximum of $\sim300$~K when within
0.8~AU ($\approx$ the softening length of the stellar potential) and reach a floor temperature of 10 K at roughly 
100~AU from the central star. {The minimum Toomre parameter in this disc
model is $Q_\mathrm{T,min}\approx4.5$, and is reached at $R=15$~AU. Thus, we
expect discs to be gravitationally stable during flyby encounters, except for those with nearly
radial orbits (not explored here) which can compress disc mass without first removing it by
action of tidal forces.}

The absence of fragmentation ensures that our runs are not bogged down by the courant condition inside a dense fragment. 
It also simplifies the analysis of our results. The main focus of this work is the
orbital evolution of the stars following an encounter. Since the torque exerted on the
stars by the disc gas depends on the mass distribution and not on the gas temperature,
the results should not depend on whether or not a disc forms secondary objects, unless
of course that owing to low temperature (low Toomre $Q$) the fragmentation is so violent 
that the entire disc is turned into a few small objects. 
 
 \subsection{Impulsively-started tidal forces}
Two relations are essential for determining the importance of tidal effects in disc-disc
interactions. The first is the ratio between the disc size $R_d$ and the pericenter
distance between the two stars $r_p$. The second is the ratio between the disc's internal angular momentum
$L_\mathrm{disc} $ and the binary system's orbital angular momentum $|\mathbf{L}_\mathrm{orb}|$. If $r_p\gg R_d$ and
$|\mathbf{L}_\mathrm{orb}|\gg L_\mathrm{disc} $, tidal effects should be negligible and the stellar orbits should approximately evolve
{ as those of point particles with mass equal to $M_*+M_d$.}

The disc internal angular momentum (measured with respect to the centre of mass of the star+disc system),
 $L_\mathrm{disc} $, can be computed analytically
assuming that the azimuthal velocity field $v_\phi$ is well approximated by the Keplerian value $v_K$. For
$p=1$,
\begin{equation}
\begin{split}
{L}_\mathrm{disc} &\approx2\pi \int_0^\infty v_K(R)\Sigma(R)  R^2 dR\\
& = \frac{\sqrt{\pi}}{2}M_d\sqrt{GM_*R_c}~~,
\end{split}
\end{equation}
which is approximately $0.835$ $M_\odot$~AU$^2$~yr$^{-1}$. Computing ${L}_\mathrm{disc} $ from
our 3D numerical models by directly summing over all cells gives
a value of ${L}_\mathrm{disc} \approx0.85$ $M_\odot$~AU$^2$~yr$^{-1}$.

Another quantity of interest is the disc's total  energy $E_\mathrm{disc} $
(consisting of the disc's total kinetic energy, the total gravitational binding energy, and the total
thermal energy). Since our simulations  are initialized with the stars on parabolic orbits (see Section~\ref{sec:orbits} below), the orbital energy of the system should be strictly zero in
the limit of zero tidal effects, i.e., when the star+disc trajectories can be accurately represented
by those of point particles.  Therefore, any discrepancy between the total energy of the binary system 
 at $t=0$ and twice the value of $E_\mathrm{disc} \approx0.24$ $M_\odot$~AU$^2$~yr$^{-2}$ provides an indication of the tidal
 forces at startup and quantifies the validity of assigning point-mass trajectories to the disc-gas in the initial conditions.

\begin{figure}
\begin{center}
\includegraphics[width=0.48\textwidth]{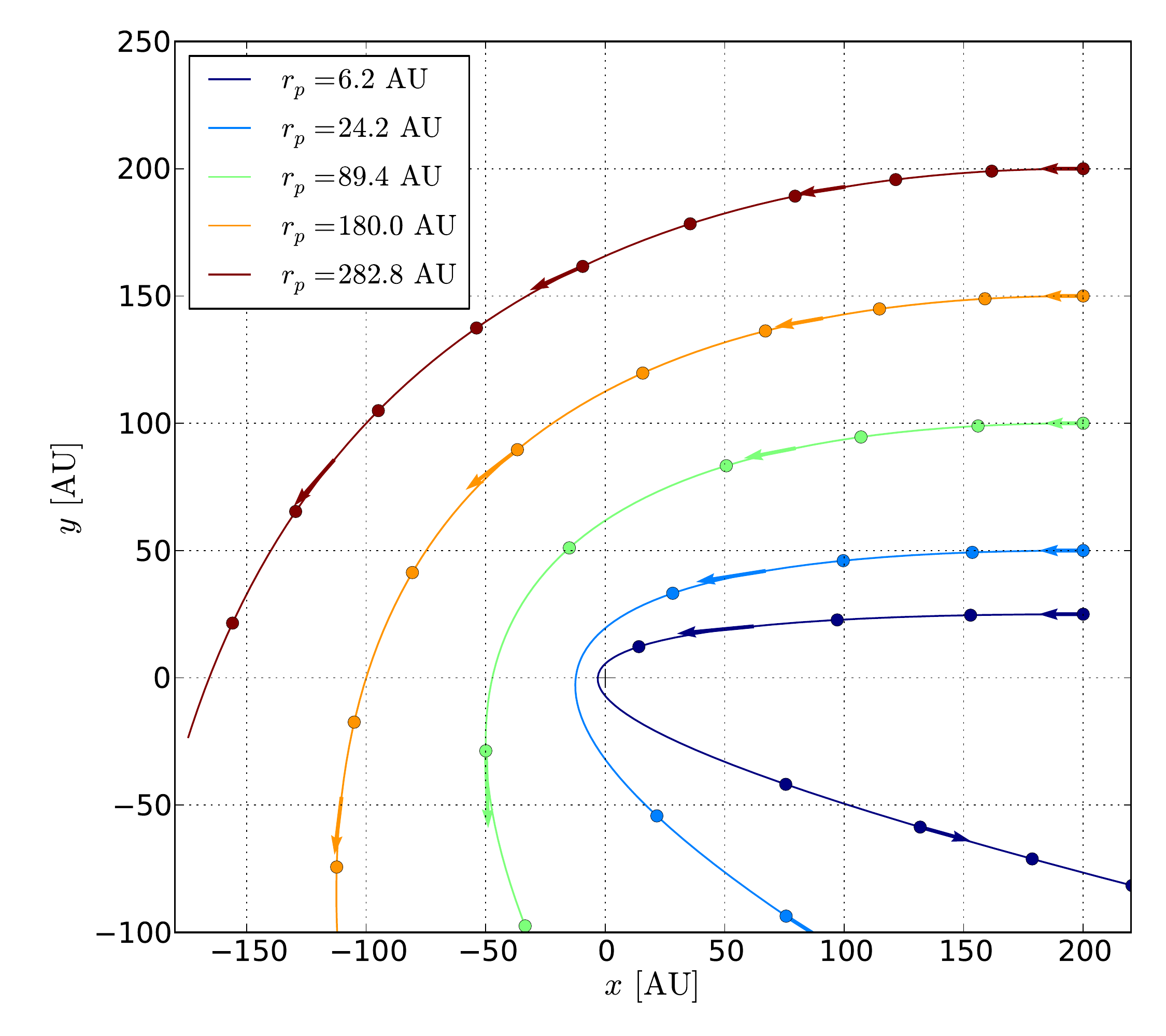}
\caption{
Orbital configurations explored in this work (for clarity, only one component
of the binary is shown). Five different parabolic
orbits are set up with five different pericenter separations, which take
values $r_p=6.2$, 24.2, 89.4, 180.0 and 282.8 AU.  Initial conditions star from
the right of the figure at $x=-200$~AU for {\it all} modeled orbits (conversely, the
binary component not shown here is started at $x=200$~AU) with velocity
along the $x$-axis. Orbital properties at $t=0$ are calculated assuming each
disc is a point particle of mass $0.5M_\odot$. The magnitude of the velocity --chosen such that the orbit
is parabolic for given an initial disc separation-- ranges from $\sim1.7$~km~s$^{-1}$
to $\sim2.1$~km~s$^{-1}$. The filled circles depict the locations for each trajectory
in 200-year intervals.
\label{fig:encounter_conf}}
\end{center}
\end{figure}
%

\begin{table*}
  \begin{center}
      \caption{Orbital parameters for parabolic encounters with different pericenter separations  \label{tab:simulations}}
    \begin{tabular}{ccccccccccl}
     \hline
     Simulation & $r_p$ & $\theta_1$ & $\phi_1$ & $\theta_2$  & $\phi_2$  &$|\mathbf{L}_\mathrm{orb}|$ & $L_{\mathrm{tot},z}$ & $M_d/M_*$  
     & disc-orbit alignment $^{a}$ & Capture (C) or \\
          & [AU] & [$^\circ$] & [$^\circ$]  & [$^\circ$]  & [$^\circ$]  & [$M_\odot$~AU$^2$~yr$^{-1}$] & [$M_\odot$~AU$^2$~yr$^{-1}$]  & &  & fly-by (FB)
\\ \hline\hline
     {\texttt{PARA1-1\_M2}} & 6.2 & 0 & 0& 0& 0&  5.53 & 7.23 & $0.05/0.45$ & P-P & FB
 \\
    {\texttt{PARA1-2\_M2}} & "  & 0 & 0& 45 & 0 & "       &  6.98  &    "           & P-P    &   C
 \\
    {\texttt{PARA1-3\_M2}} &  " & 0& 0& 90 & 0 & "        &  6.38  &    "          &  P-P   &   C
 \\
    {\texttt{PARA1-4\_M2}} &  " & 0& 0& 135 & 0& "       &  5.78  &    "           & P-R   &   C
  \\
    {\texttt{PARA1-5\_M2}} &  " & 0& 0& 180& 0& "        &  5.53  &     "         &  P-R   &   C
  \\
    {\texttt{PARA1-6\_M2}} &  " & 180& 0& 180 &0 & "   &  3.83  &    "         &  R-R   &   C
   \\
    {\texttt{PARA1-7\_M2}} &  " & 180 &0 &135 &0 & "  &   4.08  &      "        &  R-R   &   C
 \\ \hline
     {\texttt{PARA2-1\_M2}} & 24.2 & 0 & 0& 0& 0& 10.9 & 12.6 & $0.05/0.45$ & P-P    &   FB
 \\
    {\texttt{PARA2-2\_M2}} & "  & 0 & 0& 45 & 0 & "        &  12.4 & "             & P-P     &   FB
 \\
    {\texttt{PARA2-3\_M2}} &  " & 0& 0& 90 & 0 & "         &  11.8 & "           &  P-P        &   FB 
 \\
    {\texttt{PARA2-4\_M2}} &  " & 0& 0& 135 & 0&  "       &  11.2 &  "            & P-R    &   FB
  \\
    {\texttt{PARA2-5\_M2}} &  " & 0& 0& 180& 0&  "        &  10.9  &    "         &  P-R    &   C
  \\
    {\texttt{PARA2-6\_M2}} &  " & 180& 0& 180 &0 &  "   &   9.23 &   "         &  R-R    &   C
   \\
    {\texttt{PARA2-7\_M2}} &  " & 180 &0 &135 &0 &  "   &   9.48 &   "         &  R-R        &   C
 \\ \hline
     {\texttt{PARA3-1\_M2}} & 89.4 & 0 & 0& 0& 0& 21.0  & 22.7 & $0.05/0.45$ & P-P   & FB
 \\
    {\texttt{PARA3-2\_M2}} & "  & 0 & 0& 45 & 0 & "         &  22.5 &  "            & P-P   & FB
 \\
    {\texttt{PARA3-3\_M2}} & "  & 0& 0& 90 & 0 &  "         &  21.9 & "          &  P-P   & FB
 \\
    {\texttt{PARA3-4\_M2}} & "  & 0& 0& 135 & 0& "         &  21.3 &   "           & P-R   & FB
  \\
    {\texttt{PARA3-5\_M2}} &  " & 0& 0& 180& 0& "          &  21.0 &   "         &  P-R   & FB
  \\
    {\texttt{PARA3-6\_M2}} &  " & 180& 0& 180 &0 & "     &  19.3 &   "         &  R-R   & FB
   \\
    {\texttt{PARA3-7\_M2}} &  " & 180 &0 &135 &0 &  "    &  19.6 &  "        &  R-R   & FB
 \\ \hline
     {\texttt{PARA4-1\_M2}} & 180.0 & 0 & 0& 0& 0& 29.8 & 31.5 &$0.05/0.45$ & P-P   & FB
 \\
    {\texttt{PARA4-2\_M2}} & "  & 0 & 0& 45 & 0 & "          &  31.3 &  "         & P-P   & FB
 \\
    {\texttt{PARA4-3\_M2}} & "  & 0& 0& 90 & 0 & "           &  30.7 &   "           &  P-P   & FB
 \\
    {\texttt{PARA4-4\_M2}} & "   & 0& 0& 135 & 0&  "        &   30.1 &  "             & P-R    & FB
  \\
    {\texttt{PARA4-5\_M2}} &  " & 0& 0& 180& 0&  "          &   29.8 &   "         &  P-R    & FB
  \\
    {\texttt{PARA4-6\_M2}} &  " & 180& 0& 180 &0 & "      &   28.1 & "          &  R-R   & FB
   \\
    {\texttt{PARA4-7\_M2}} &  " & 180 &0 &135 &0 & "      &   28.4 & "         &  R-R   & FB
     \\ \hline
     {\texttt{PARA5-1\_M2}} & 282.8 & 0 & 0& 0& 0& 37.4 & 39.1 &$0.05/0.45$ & P-P   & FB
 \\
    {\texttt{PARA5-2\_M2}} &  " & 0 & 0& 45 & 0 &  "         & 38.8  & "             & P-P   & FB
 \\
    {\texttt{PARA5-3\_M2}} &  " & 0& 0& 90 & 0 &  "          & 38.2  & "           &  P-P   & FB
 \\
    {\texttt{PARA5-4\_M2}} &  " & 0& 0& 135 & 0&  "         & 37.6  &   "            & P-R   & FB
  \\
    {\texttt{PARA5-5\_M2}} &  " & 0& 0& 180& 0& "           & 37.4  &   "         &  P-R   & FB
  \\
    {\texttt{PARA5-6\_M2}} &  " & 180& 0& 180 &0 & "      & 35.7   &  "         &  R-R   & FB
   \\
    {\texttt{PARA5-7\_M2}} &  " & 180 &0 &135 &0 & "      & 35.9   &   "        &  R-R   & FB
 \\ \hline
    \end{tabular}
     $^{a}$ disc-orbit alignment is classified into prograde-prograde (P-P), prograde-retrograde (P-R) or
 retrograde-retrograde (R-R) based on the orientation angles $\theta_i$ with $i=1,2$  
 (prograde if $\theta_i\leq90^\circ$ and retrograde if $\theta_i>90^\circ$).\hspace*{\fill}
  \end{center}
 \end{table*}

\subsection{Orbital configurations}\label{sec:orbits}
Two identical copies of the fiducial disc presented in Section~\ref{sec:disc_models}
are used to set up a parabolic encounter. Consequently, the orbital energy of the binary is $E_\mathrm{orb}=0$.
{The orbital plane of the encounter coincides with $x$-$y$ plane and the star-disc bulk initial velocities are
directed along the $x$-axis (the discs directed toward each other)}. Therefore,
the only free parameter is the pericenter separation $r_p$. Figure~\ref{fig:encounter_conf} shows
the initial trajectories of one of the discs for five different values of $r_p$: 6.2, 24.2, 89.4, 180.0
and 282.8~AU. We call these different parabolic orbit configurations \texttt{`PARA1'}, \texttt{`PARA2'},
\texttt{`PARA3'}, \texttt{`PARA4'} and \texttt{`PARA5'} respectively.
In addition, we vary the orientation of the discs with respect to the orbital angular
momentum vector (angles $\theta_1$ and $\theta_2$). We also vary the disc orientations in seven
different configurations, which are labeled accordingly by appending a number to the orbital
label, e.g., \texttt{`PARA1-1'}, \texttt{`PARA1-2'}, etc. Table~\ref{tab:simulations} shows the
main set of simulations and their respective orbital and orientation parameters. Each
orbital configuration (set by the value of $r_p$) contains seven variants, which correspond
to different combinations of the angles $\theta_1$ and $\theta_2$ (same notation as \citealp{she10}).
The azimuthal orientation of the discs (angles $\phi_1$ and $\phi_2$) is not changed. 

The orbital angular momentum in the two-body problem is 
$\mathbf{L}_\mathrm{orb}=m\sqrt{\mu r_p (1+e)}\,\hat{\mathbf{z}}$
where $m=M_1M_2/(M_1+M_2)=0.25M_\odot$ is the reduced mass and $\mu=GM_\mathrm{tot}$
where the total mass is $M_\mathrm{tot}=1M_\odot$. The seventh column in Table~\ref{tab:simulations}
shows the orbital angular momentum of each simulation according to the chosen value of $r_p$. The
eighth column shows the expected value of the $z$-component of the {\it total} angular momentum $L_{\mathrm{tot},z}$, taking
into account the contribution from the disc internal angular momentum; i.e., at time $t=0$:
\begin{equation}\label{eq:total_angmom}
L_{\mathrm{tot},z}= |\mathbf{L}_\mathrm{orb}|+ L_{\mathrm{disc} ,1}\cos\theta_1 + L_{\mathrm{disc} ,2}\cos\theta_2~~,
\end{equation}
where $L_{\mathrm{disc} ,1}=L_{\mathrm{disc} ,2}\approx0.85$~AU$^2$~yr$^{-1}$ (Section~\ref{sec:disc_models}). A comparison between $|\mathbf{L}_\mathrm{orb}|$ and
$L_{\mathrm{tot},z}$ shows that the total angular momentum can be changed significantly by simply changing
the orientation of the discs. From these quantities, we can estimate that the simulation subsets \texttt{`PARA1'} and
\texttt{`PARA2'} should show a greater degree of redistribution of angular momentum between the gas and the stars 
(and produce capture) as well as a significant dependence of the simulation outcome on the orientation of the discs.
Our initial conditions satisfy Equation~(\ref{eq:total_angmom}) within less
than a few percent, indicating that the superposition of two stationary, isolated disc models into a 
self-interacting binary system is reasonably adequate at the values of the initial separation $D$ 
that we have chosen. We have noticed
that the angular momentum error seeded on startup is slightly higher for the larger pericenter simulations. Although
the discs in these simulations were started far apart (Figure~\ref{fig:encounter_conf}) precisely to avoid these problems,
it is worth pointing out that the angular momentum of the system grows faster with $r_p$ than with $D$. For example,
the ratio in $D$ for the orbital configurations \texttt{`PARA5'} ($r_p=282.8$) and \texttt{`PARA1'} ($r_p=6.2$) is $\sim1.4$,
while the ratio in $|\mathbf{L}_\mathrm{orb}|$ for the same configurations is $\sim6.8$. As a consequence, the errors in
setting up the orbit are fractionally larger, albeit still small, for our wider orbits. Although the impulsive forces implied
by this fractional error are still negligible for the dynamics of the encounter, we foresee some small complications arising
from not setting up a configuration that is not sufficiently distant.

Similarly, the consistency between the measured energy at the zeroth snapshot and the expected energy
derived from the superposition of two isolated disc models provides another measure of whether
the initial conditions are quiet, or if the discs are so close that the tidal forces are imparted impulsively at the beginning of the simulation.
 The total energy of the system is $E_\mathrm{tot}=2 E_\mathrm{disc} + K_\mathrm{orb}+V_\mathrm{orb}$,
but as we have mentioned above, $ K_\mathrm{orb}+V_\mathrm{orb}=E_\mathrm{orb}\approx0$. Therefore,
the total energy of the system, {\it regardless} of the pericenter distance,  should be just twice the energy of the
 individual disc model: $E_\mathrm{tot}\approx-0.48$~AU$^2$~yr$^{-2}$. 
 Just as  with $L_\mathrm{tot}$, we find deviations of
the order of $1-2\%$, indicating that some tidal forces are acting on the discs at $t=0$.

\section{Results}~\label{sec:results}
%

\subsection{Encounter morphology}

\begin{figure*}
\vspace{-0.08in}
\centering
\includegraphics[width=0.82\textwidth]{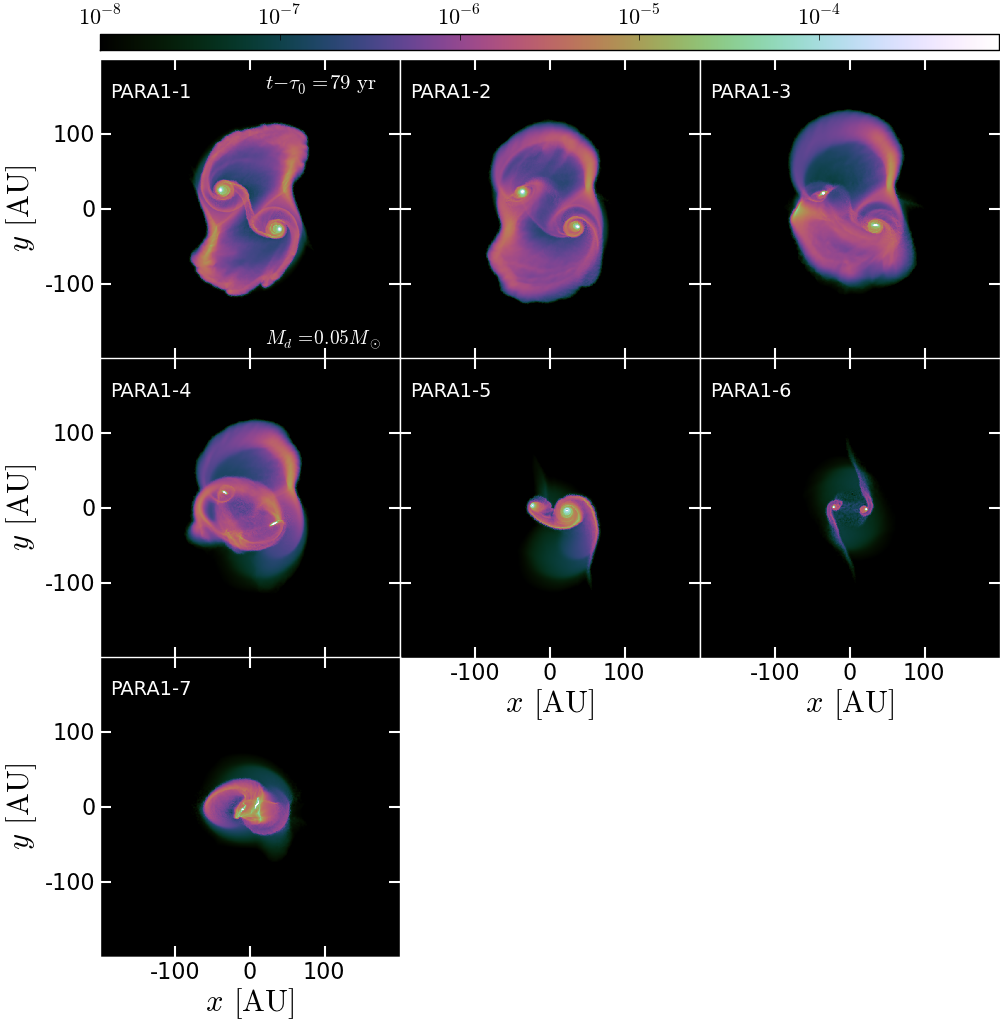}
\vspace{-0.15in}
\caption{Simulation output for the orbital set \texttt{`PARA1'} ($r_p=6.2$) a short time after pericenter
passage ($t-\tau_0=79$~yr), which corresponds to a simulation time of 700 yr. Six out of these
seven simulations show orbital capture before the end of the integration (5000 years), meaning
that the stars came back for at least one more pericenter passage (see text).
Each frame shows the projected density in units of $M_\odot$~AU$^{-2}$ (the conversion
factor to g~cm$^{-2}$ is $8.88\times10^6$). All images are generated by integrating the three-dimensional
density field along one direction following the full Voronoi mesh. 
\label{fig:encounter_PARA1}}
\vspace{-0.08in}
\end{figure*}

\begin{figure*}
\vspace{-0.08in}
\centering
\includegraphics[width=0.82\textwidth]{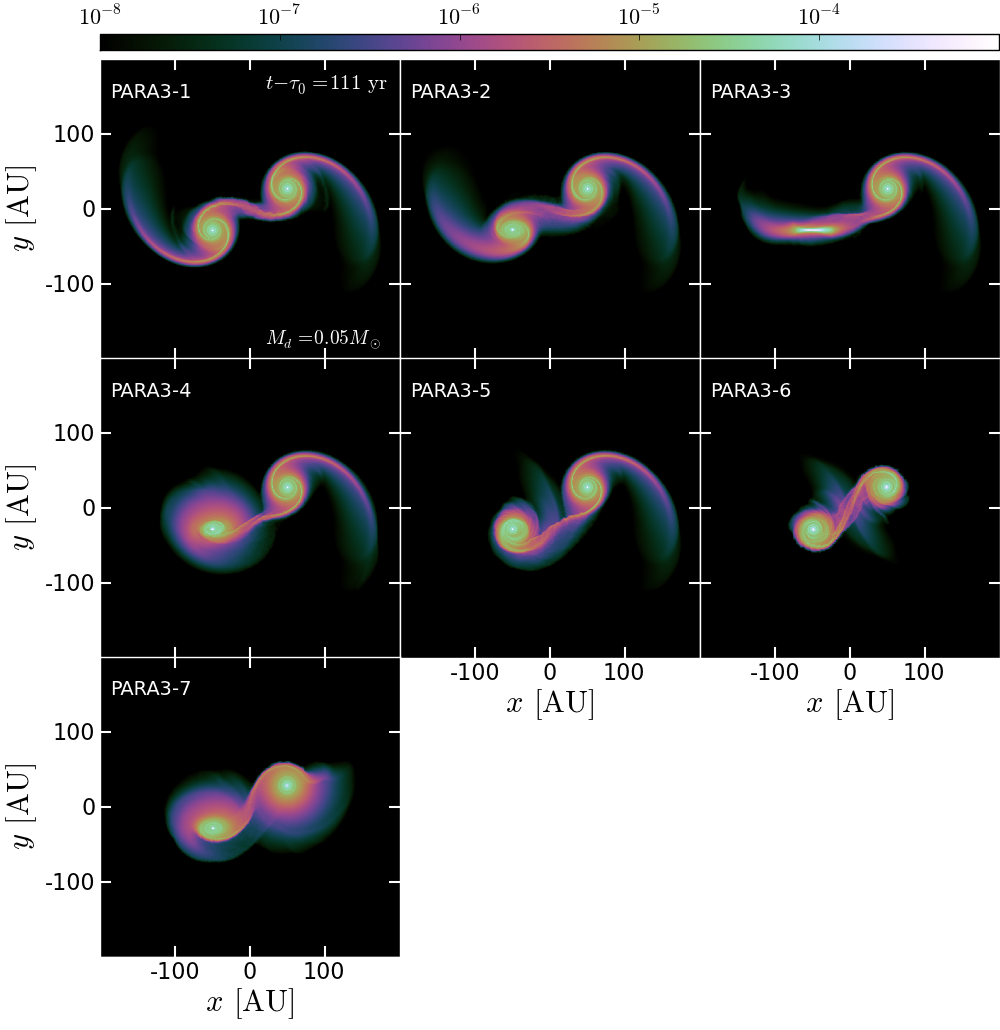}
\vspace{-0.15in}
\caption{Simulation output for the orbital set \texttt{`PARA3'} a short time after pericenter
passage ($t-\tau_0=111$~yr), which corresponds to a simulation time of 1000 yr. See description
of Figure~\ref{fig:encounter_PARA1}.
\label{fig:encounter_PARA3}}
\vspace{-0.08in}
\end{figure*}

\begin{figure*}
\vspace{-0.18in}
\centering
\includegraphics[width=0.82\textwidth]{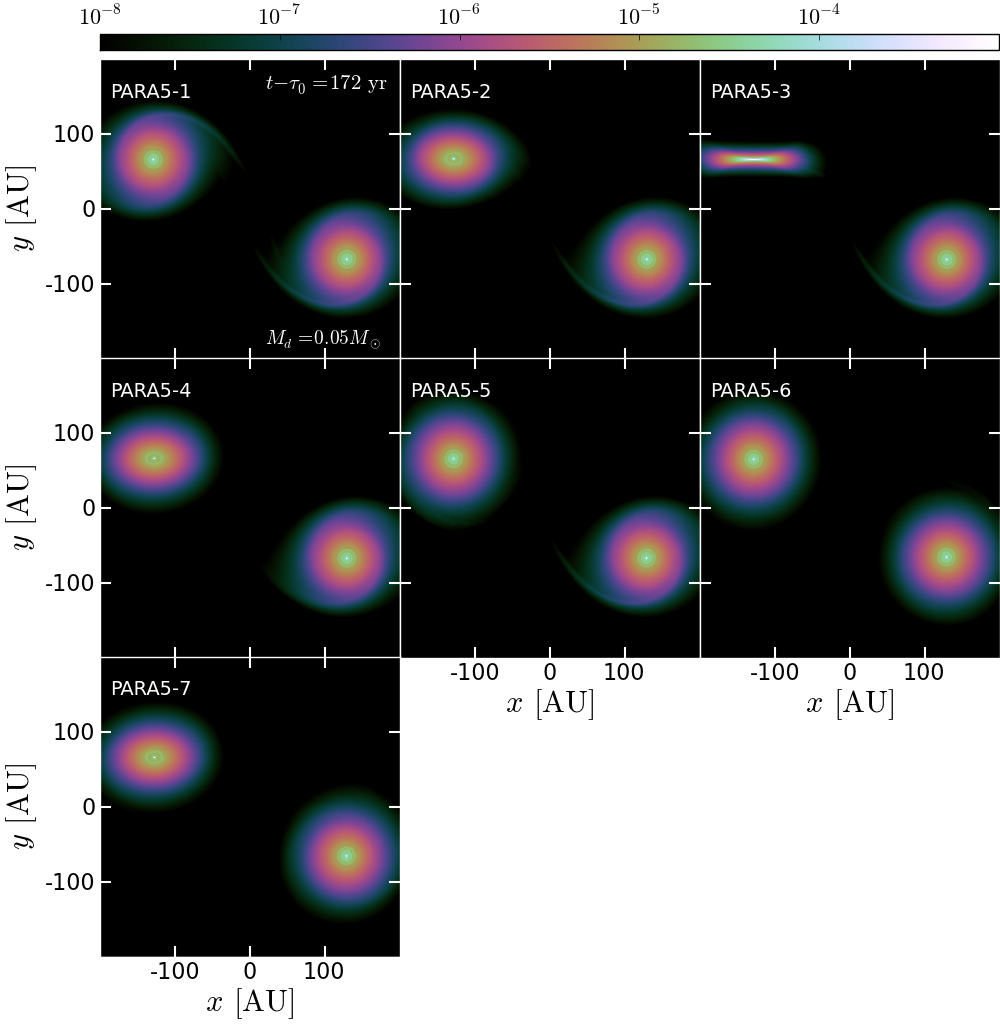}
\vspace{-0.17in}
\caption{Simulation output for the orbital set \texttt{`PARA5'} a short time after pericenter
passage ($t-\tau_0=172$~yr), which corresponds to a simulation time of 1600 yr. See description
of Figure~\ref{fig:encounter_PARA1}.
\label{fig:encounter_PARA5}}
\vspace{-0.17in}
\end{figure*}

Figures~\ref{fig:encounter_PARA1}-\ref{fig:encounter_PARA5} show projected
density images of 21 out of the 35 simulations listed in Table~\ref{tab:simulations}. 
These figures show rendered snapshots at some time after pericenter passage for
simulation subsets \texttt{`PARA1'}, \texttt{`PARA2'} and \texttt{`PARA5'}  
(Table~\ref{tab:simulations}). 

Simulation sets \texttt{`PARA1'}  and \texttt{`PARA2'}  have pericenter distances
of $r_p=6.2$ and $24.2$~AU respectively, and therefore we expect the greatest
disruption  to the gas discs in these simulation sets due to tidal effects
but also due to direct shock-induced truncation. The disc models have
characteristic radii of $R_c=20$~AU and outer radii of $R_d=60$~AU (Section~\ref{sec:disc_models}),
and thus the discs are expected to collide directly (i.e., $r_p/2< R_d$) for configurations
\texttt{`PARA1'}, \texttt{`PARA2'} and \texttt{`PARA3'}.  However, for \texttt{`PARA3'}, the
enclosed mass at $R=r_p/2=44.7$ is  $89\%$ of the disc total mass, and thus, despite
the evident disc truncation seen in Figure~\ref{fig:encounter_PARA3} (note that the projected
density range spans nearly 5 orders of magnitude), this encounter should have little effect on the stellar orbits.


Indeed, two different
regimes can be clearly distinguished from the evolution of the stellar orbits. 
Figure~\ref{fig:encounter_PARA1}
shows significant disruption of the gas distribution, 
as well as changes in the orbit of the stars. In contrast, figure~\ref{fig:encounter_PARA3} shows
some disc truncation and strong tidal features, but the disc centres (approximately
the position of the stars) show no distinguishable variation from frame to frame, i.e., the
orientation of the discs bears little importance. The response of the 
orbits depends weakly on the extended mass distribution thus the evolution
can be reasonable approximated by two point masses.
Figure~\ref{fig:encounter_PARA5} shows little
to no modification in the disc surface density besides the excitation of $m=2$
spiral arms, which are characteristic of tidal encounters \citep[e.g.][]{bin08,don10}.
Evidently, the steepness of the tidal force with distance can explain the rapid change
in output with $r_p$ (the calculations of \citealp{ost94}
show an exponential dependence of the energy change in $(r_p/R_d)^{3/2}$).
 
Another indication of the rapidly decreasing influence of the disc is the presence of
some tidal features. {We can see the same tidal
features arise in disc $\#1$ in the first five panels of Figure~\ref{fig:encounter_PARA3}
and~\ref{fig:encounter_PARA5}  (morphological similarities are
harder to find in Figure~\ref{fig:encounter_PARA1}). These features arise
independently of the orientation and shape
of disc $\#2$ (the orientation of disc $\#1$ is unchanged for the first five runs
of each simulation subset in Table~\ref{tab:simulations}). This
 is an indication that, to first order,} the tidal features depend on the monopole component of
the companion's potential (strongly dominated by the central star), even as significant mass stripping
has taken place due to the physical collision of the discs.

\subsection{Stellar orbits}

Figure~\ref{fig:separation_PARA1} shows the inter-star separation for the most disruptive set of orbits (\texttt{`PARA1'}).
At pericenter passage, all orbits suffer a significant energy loss that shrinks the semi-major axis compared to the initial parabolic trajectory. Before pericenter (i.e., $t<$621~yrs) all runs
follow the analytic trajectory closely, and reach pericenter at the same time. After pericenter, the orbital evolution
varies dramatically among these configurations. In six of the runs, there are at least 
three additional close pericenter passages 
and over $\sim50$ additional ones in the most disruptive configuration. Based on these additional
separation minima we categorize these systems as ``captured."
The seventh configuration (\texttt{`PARA1-1'}) hints that 
the stellar separation has reached apocenter at the end of the simulation, and that it should go back for a second passage
at around $t=7000$~yr. Consequently, all simulations in the set \texttt{`PARA1'} show sufficient energy loss to be considered
bound after first passage. 

The different outcomes of the \texttt{`PARA1'} simulations are determined by the relative orientation of the discs. Each
curve in Figure~\ref{fig:separation_PARA1} is labeled according to a normalized
$z$-spin value $S_z\equiv\cos\theta_1+\cos\theta_2$. Prograde-prograde encounters like \texttt{`PARA1-1'},
\texttt{`PARA1-2'} and \texttt{`PARA1-3'} contain a larger amount of angular momentum than the prograde-retrograde
and the retrograde-retrograde encounters. Configurations that include one or two discs in retrograde orientation
will result in different torques on to the stars since the initial response of the gas to first passage will be different. Retrograde
orientations do not contain orbital resonances \citep[e.g.][]{too72,don10}, and extended spiral arms are not formed as a result.
The parabolic orbits in this models are chosen to model the first ever encounter between two objects (hyperbolic encounters
were not considered in this work), and thus we do not expect a correlation between pericenter separation and mutual
disc inclination. This is most likely not true for bound orbits, however, since a common formation scenario for close binary stars
would suggest that these are born with nearly aligned discs as they are with nearly aligned spins \citep[see][]{hal94}.

The orbital decay is almost entirely determined by the energy loss at first passage. Although the amplitude
of the separation curve is observed to decay slowly in time, most of the dissipation happens at once when the discs
first meet. This is not surprising given that configuration \texttt{`PARA1'} has a pericenter distance ($r_p=6.2$) that is significantly
smaller than the sum of the disc radii ($2\times R_d=120$~AU). Thus, the first encounter violently truncates the disc on very short
timescales, potentially reducing the mass of the disc by a factor of $\sim4$ (from Equation~\ref{eq:enclosed_mass}, 
$M_d(<6.2\,\mathrm{AU})\approx M_d/4$), after which the tidal interaction goes back into a linear regime and the
orbit evolves more slowly. 

Figure~\ref{fig:separations} shows the evolution of stellar separation for the remainder of the simulations in our study,
grouped by orbital configuration. The likelihood of capture decreases very rapidly with
pericenter separation. Only three out of seven simulations in the \texttt{`PARA2'} show additional pericenter passages
(although all of them show  substantial orbital energy loss). The runs in set  \texttt{`PARA3'} show a much weaker effect;
although, as before, the change in separation increases when $S_z$ is decreased.  The stars should not be expected
to interact again for another few 10,000~yrs, and after reaching separations of a few to several thousand AU, meaning
that these systems are not true ``binaries". In the case of \texttt{`PARA4'} and \texttt{`PARA5'}, the
interaction appears extremely weak, since stellar separations remain on their original parabolic trajectories
with variations of the order of $1\%$ toward the end of the simulation. These variations are attributable to other effects besides
tidal interactions (e.g., numerical accuracy, lack of strict conservation of angular momentum, torque onto the stars by gas accretion, etc).
Although the tidal response of the disc is clear in these last two examples, the long term effect on the orbits of the stars is
negligible.

\subsection{Orbital evolution}
Although the stellar separation is informative, it does not describe completely the
orbital evolution. In order to analyze the orbital
evolution of each star+disc system, we need to assign them meaningful osculating elements. However,
identifying which gas belongs to a disc and which is simply surrounding material is not a trivial task.
Although group finding algorithms abound in the literature, these are designed for heavily clustered distributions
in space (this is the basis for halo-finding algorithms in cosmology; e.g., \citealp{dav85}). Although 
spatial density-based clustering algorithms might have little trouble identifying discs in configurations
as in Figure~\ref{fig:encounter_PARA5}, disordered gas distributions like that of Figure~\ref{fig:encounter_PARA1}
likely present a challenge for automized searches. For simplicity, we will consider as disc material all
cells lying within each star's Roche lobe (in this equal-mass example, delimited roughly by the midpoint between the two stars).
Then, we proceed to identify all the mass within the chosen region, calculating its centre of mass and the centre of mass
velocity. With these quantities, we define a classical two-body problem and calculate the orbital elements
for each snapshot. Since over short timescales the dynamics should be dominated by the two stars (they contain
$90\%$ of the mass of the system), we expect this approximation to be a good first order indicator of the
orbital evolution of the system.

\begin{figure}
\begin{center}
\includegraphics[width=0.45\textwidth]{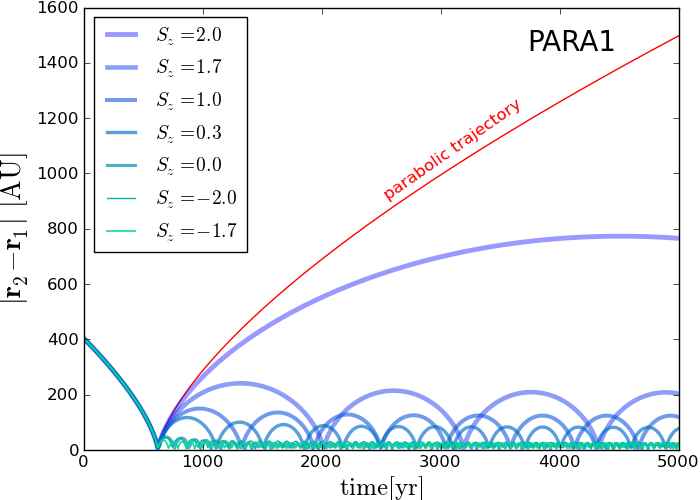}
\vspace{-0.1in}
\caption{Time evolution of stellar separation in the \texttt{`PARA1'} orbital
configuration (seven distinct runs). Lines are labeled according to ``normalized total disc spin" in the $z$-direction, 
defined as $S_z=\cos\theta_1+\cos\theta_2$. If $S_z=2$, both discs have spin angular
momenta aligned with $\mathbf{L}_\mathrm{orb}$, while a  value of $-2$ means that both spins are
antiparallel with $\mathbf{L}_\mathrm{orb}$. The simulation results show that the exactly parabolic intial
trajectory (red line) is followed very closely by the stars until pericenter passage. 
\label{fig:separation_PARA1}}
\end{center}
\end{figure}
\begin{figure*}
\includegraphics[width=0.43\textwidth]{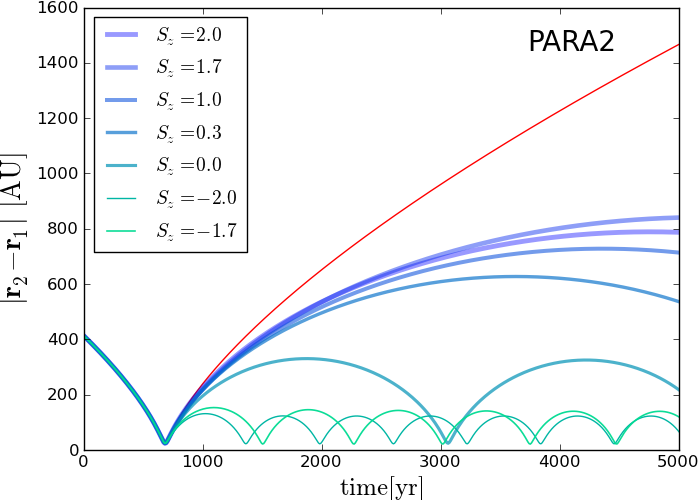}
\includegraphics[width=0.43\textwidth]{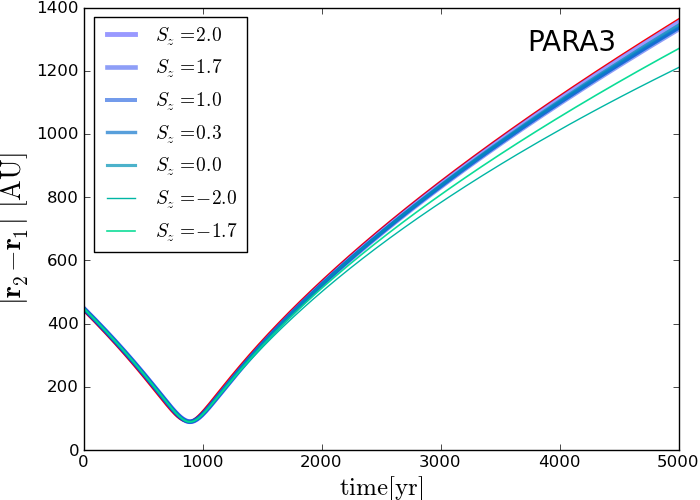}
\includegraphics[width=0.43\textwidth]{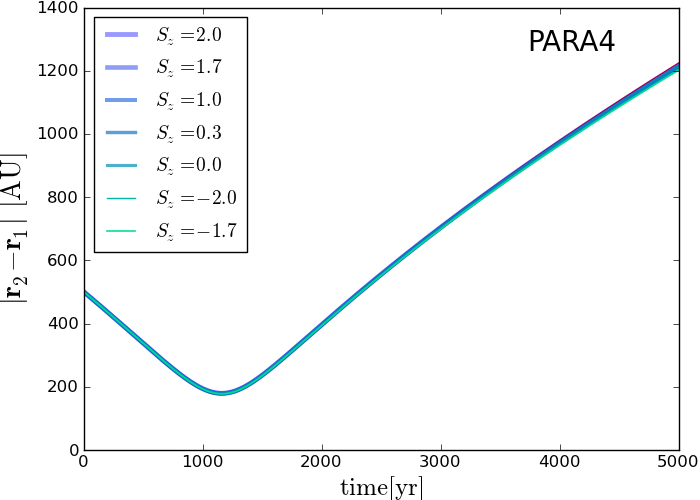}
\includegraphics[width=0.43\textwidth]{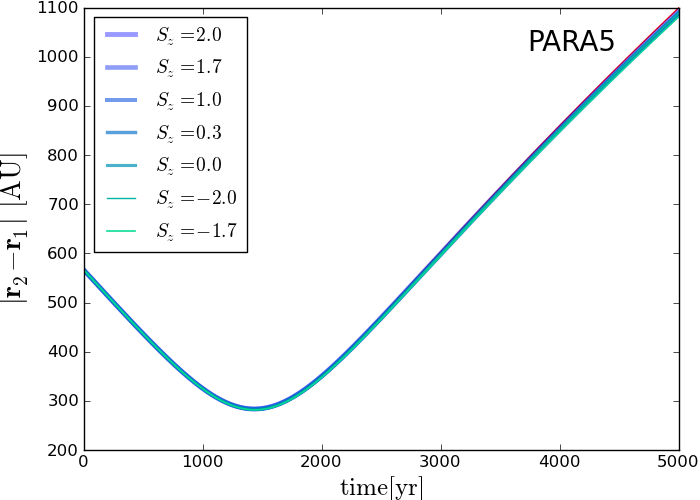}
\caption{Same as Figure~\ref{fig:separation_PARA1}, but now for the 
 \texttt{`PARA2'},  \texttt{`PARA3'}, \texttt{`PARA4'} and \texttt{`PARA5'} configurations. As before, 
 the color of the curves is chosen according to the combined disc spin in each configuration
\label{fig:separations}}
\end{figure*}

\begin{figure*}
\centering
\includegraphics[width=0.85\textwidth]{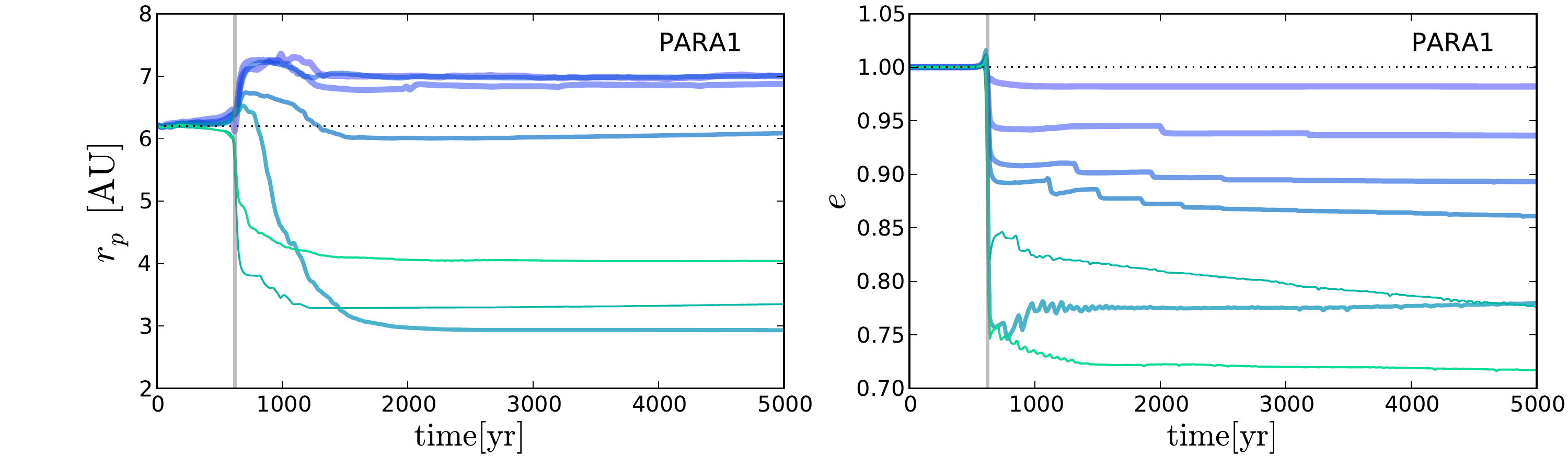}
\includegraphics[width=0.85\textwidth]{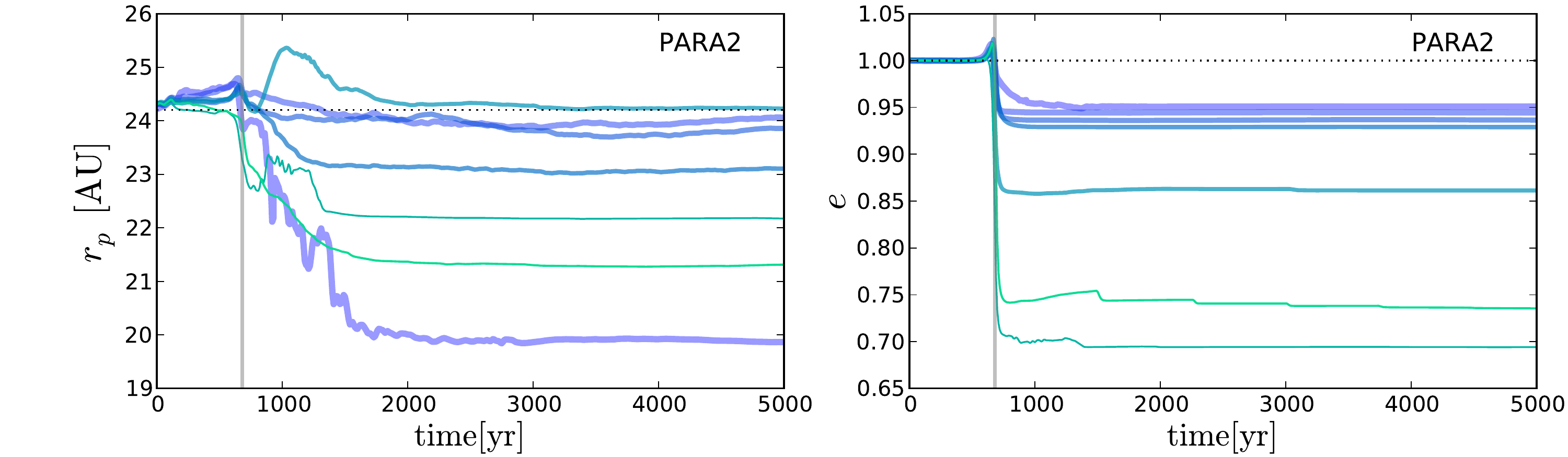}
\includegraphics[width=0.85\textwidth]{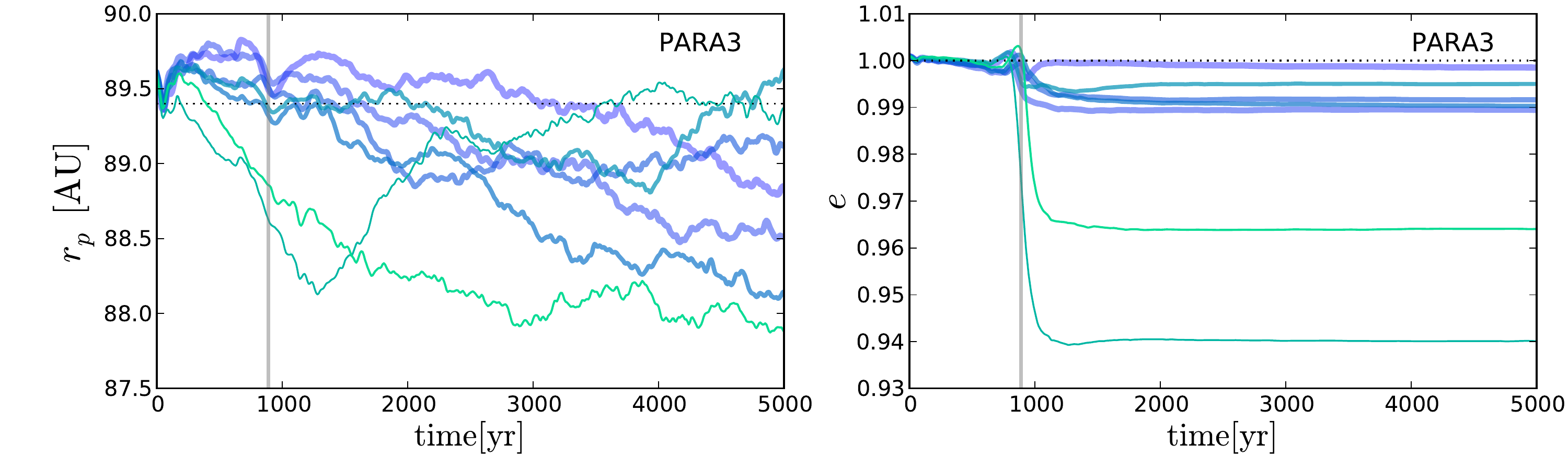}
\caption{Time evolution of the orbital elements (pericenter separation $r_p$ and eccentricity $e$)
for \texttt{`PARA1-PARA3'} from Table~\ref{tab:simulations} using the same color scheme as in Figure~\ref{fig:separations}.
 \texttt{`PARA4-5'} are omitted due to similarity with \texttt{`PARA3'} as described in the text.
The vertical gray lines mark the expected time of pericenter $\tau_0$ based on the initial parabolic orbit for any given orbital
configuration. The horizontal dotted line represents the initial value of $r_p$ (Table~\ref{tab:simulations}) and eccentricity $e$
(=1 for all orbits). 
\label{fig:elements_results}}
\end{figure*}

Figure~\ref{fig:elements_results} shows the orbital elements calculated in the manner described
above for a subset of the simulations. The pericenter time series show a markedly different behaviour
between simulation sets \texttt{`PARA1'} and  \texttt{`PARA2'} with respect to sets \texttt{`PARA3'},
\texttt{`PARA4'} and \texttt{`PARA5'}. The former group shows substantial changes in $r_p$
after pericenter (up to $50\%$ in the case of \texttt{`PARA1'}), while the latter group
shows a variability that could be consistent with a random walk behaviour. Smoother over
timescales of $\sim1000$, sets \texttt{`PARA3'},
\texttt{`PARA4'} and \texttt{`PARA5'}  show little to no consistent trend in $r_p$ as
a function of time. The total $\Delta r_p$  ranges from $1\%-2\%$ in the case of \texttt{`PARA3'},
to $\sim0.1\%$ in the case of  \texttt{`PARA5'}. The seemingly stochastic behaviour of $r_p$ in this
regimes leads us to conclude that the evolution is dominated by numerical noise or stochastic accretion
due to our sink particle scheme and that the ``true" small variability of $r_p$ is buried (\texttt{`PARA4'} and \texttt{`PARA5'})
or partially buried (\texttt{`PARA3'}) under noise.

The evolution of eccentricity shows a significant deviation from unity for sets \texttt{`PARA1'} (up to $30\%$),
\texttt{`PARA2'} (up to $30\%$) and \texttt{`PARA3'} (up to $6\%$). Even \texttt{`PARA4'} and \texttt{`PARA5'}
(not shown),
although again with a component of stochasticity, show a clear overall trend of decreasing eccentricity
that flattens out toward the end of the simulation. The fact that the eccentricity reaches a finite value toward
the end indicates that, despite the evident noise contamination, this loss of energy is real; furthermore, \texttt{`PARA5'}
flattens out later than \texttt{`PARA4'}, consistent with the fact that the pericenter timescale of \texttt{`PARA5'}
is longer and therefore the tidal interaction is expected to be spread over a longer time interval.

An interesting outcome of the pericenter evolution of \texttt{`PARA1'} is that $r_p$ grows after pericenter for positive
values of $S_z$ but decreases for $S_z\leq0$. Therefore, although Figure~\ref{fig:separation_PARA1}
already hints at a loss of orbital energy (confirmed by the drop in $e$ below $1$ right after pericenter) the pericenter does not necessarily
shrink. In principle, this effect can shield the discs from undergoing a second disruptive  pericenter passage of similar proximity
to the first, now that the minimum distance has been increased. Conversely, those simulations with the most negative
values of $S_z$ show a decrease in the magnitude of pericenter distance after the first passage.
These simulations (\texttt{`PARA1-6'} and \texttt{`PARA1-7'}) show several subsequent encounters, as if undergoing
a runaway process in which each encounter facilitates the following one at an even smaller separation.
This process could only stop once the gravitational softening lengths of the stars overlap (thus introducing an artificial ``pressure"),
or if the dispersal of the disc -- via truncation or accretion -- has made the tidal effects insignificant. Indeed all
 negative spin simulations in \texttt{`PARA1'}
end up stabilizing in $r_p$, although they do so at a time considerably longer than the timescale associated with pericenter
passage.

\begin{figure*}
\vspace{-0.08in}
\centering
\includegraphics[width=0.82\textwidth]{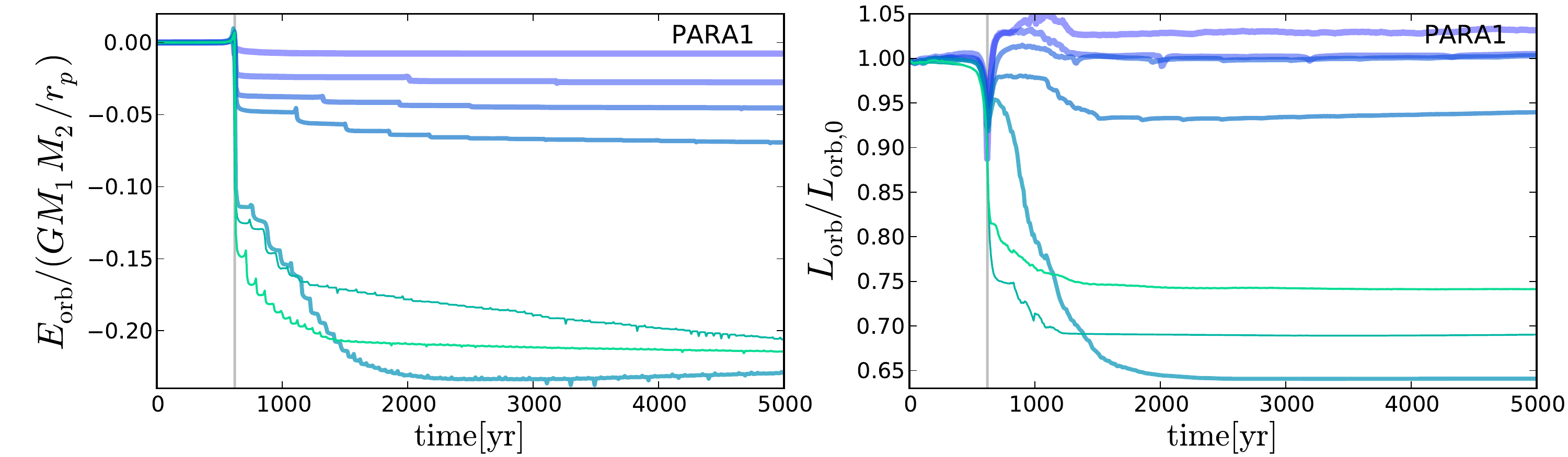}
\includegraphics[width=0.82\textwidth]{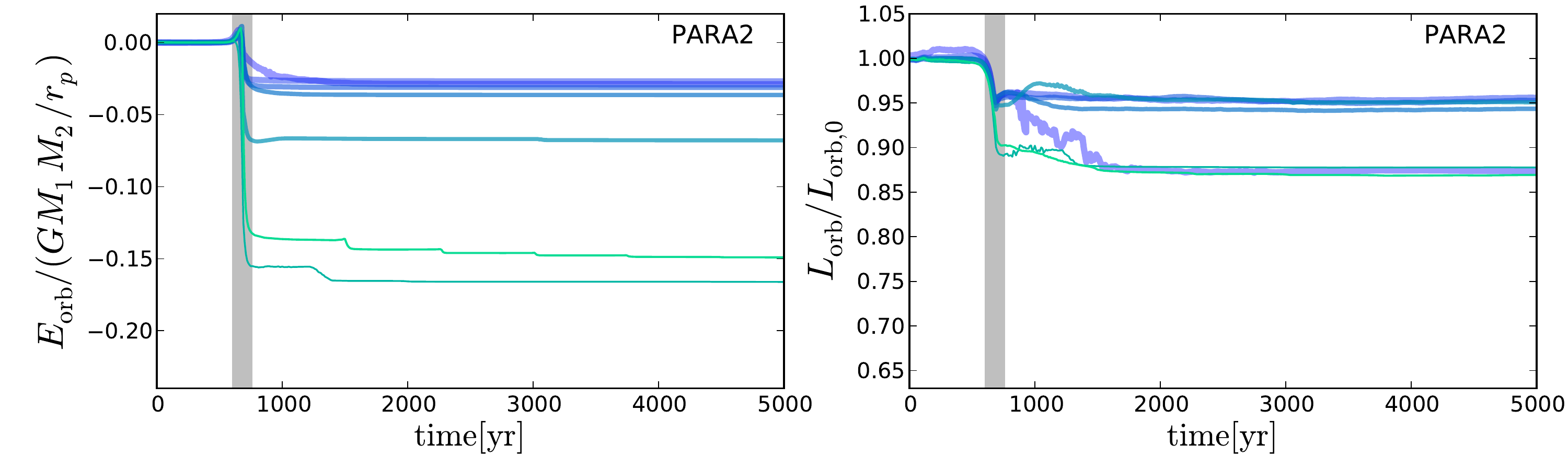}
\includegraphics[width=0.82\textwidth]{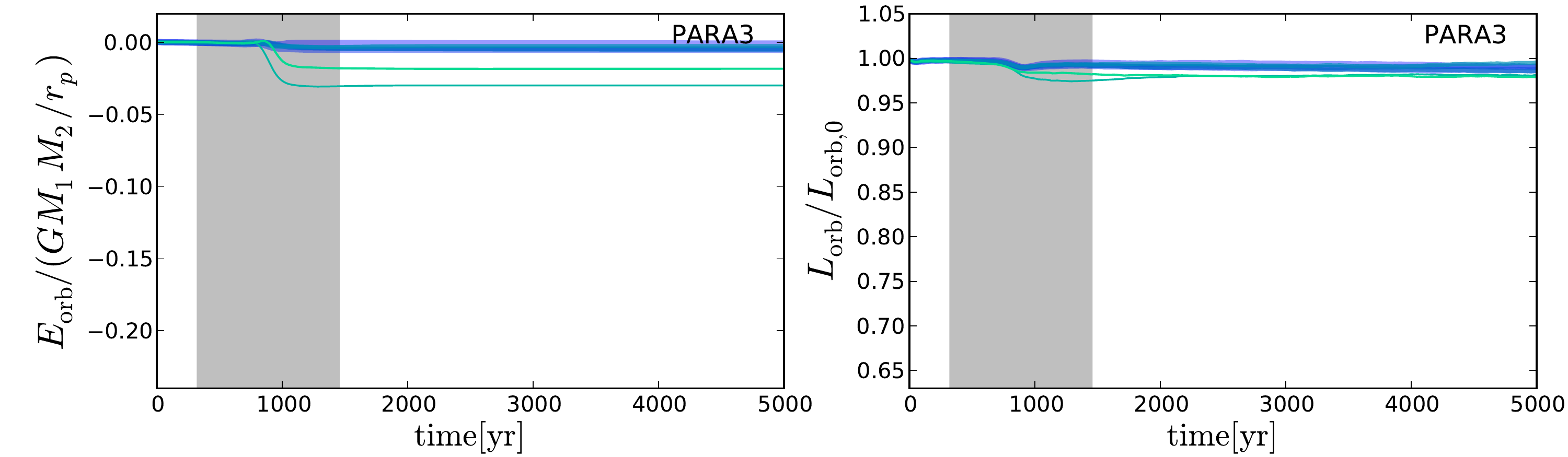}
\includegraphics[width=0.82\textwidth]{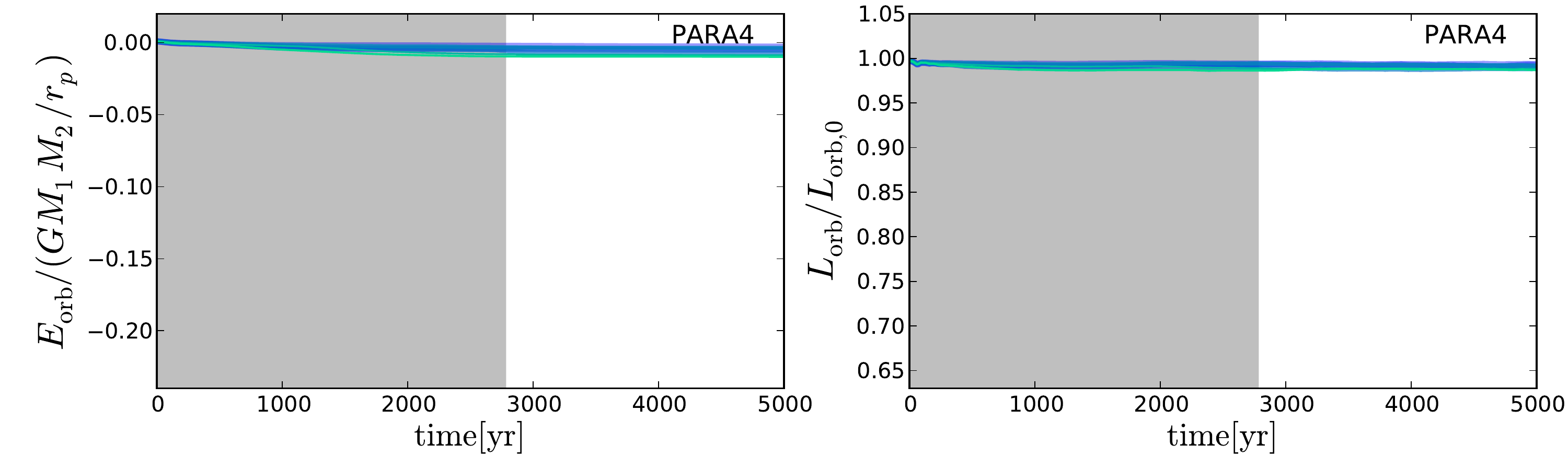}
\includegraphics[width=0.82\textwidth]{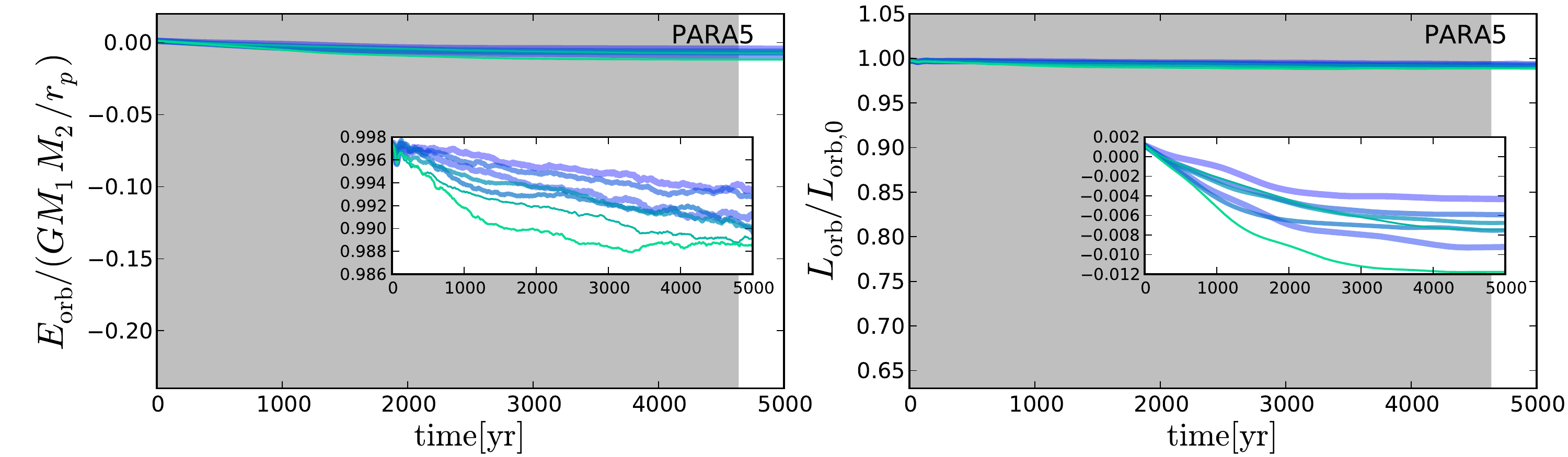}
\vspace{-0.08in}
\caption{Orbital energy $E_\mathrm{orb}$ (left column) normalized by the reference value
$GM_1M_2/r_p$ and orbital angular momentum $|\mathbf{L}_\mathrm{orb}|$
normalized by its initial value $L_{\mathrm{orb},0}=\tfrac{1}{4}$ for the 35 simulations of Table~\ref{tab:simulations}
using the same color scheme as in Figure~\ref{fig:separations}.
Each panel groups a subset of simulations
of identical pericenter distance. The coordinate range in the $y$-axes is the same for all orbital configurations
to highlight the dramatic difference in orbital evolution as $r_p$ increases. The inset in the last row (orbital
configuration \texttt{`PARA5'}) contains a zoomed in region showing that the change in  $E_\mathrm{orb}$
and  $|\mathbf{L}_\mathrm{orb}|$ is less than $1\%$. The shaded region covers the ``interaction period" centered
around pericenter within which most of the energy and angular momentum exchange between the two
discs takes place. \texttt{'PARA3'} and \texttt{`PARA4'} (not displayed) show variability of order $5\%$. 
\label{fig:energy_results}}
\end{figure*}

Another feature observed in the first three orbital configurations and absent in \texttt{`PARA4'} and \texttt{`PARA5'} is the
sharp increase in eccentricity right before pericenter. Note that for those orbits that were classified under ``orbital capture"
(Table~\ref{tab:simulations}), every subsequent pericenter passage is preceded by smaller glitches in eccentricity.
Technically, this means that right before the orbit becomes elliptical
it actually behaves briefly as a hyperbolic orbit. One must bear in mind that these orbital elements are proxies for what
is actually happening with the (at time ill-defined) discs during the encounter, and that these values of $r_p$ and $e$ might not have much
physical meaning when the gas is being entirely dispersed by a very violent interaction. Indeed, an important transition
 when going from \texttt{`PARA3'} to \texttt{`PARA4'} is that in the former case the discs actually come into contact,
 while in the latter there is no direct gas collision. Therefore, glitches in eccentricity observed right before a close encounter
 might be an exclusive outcome of disc-disc interactions mediated by shocks. Alternatively, the glitch in eccentricity
 for \texttt{`PARA4'} and \texttt{`PARA5'} is either too mild to be detected above the noise fluctuations or it simply
 should have taken place before the start of the simulation (see Section~\ref{sec:orbital_energy}), which would
 explain why as opposed to the other orbital configurations, eccentricity in  \texttt{`PARA4'} and \texttt{`PARA5'}
 is decreasing roughly monotonically in time since the beginning of the runs.

\subsection{Orbital energy and angular momentum}\label{sec:orbital_energy}
To track the evolution of the binary orbits, we calculate the orbital energy  $E_\mathrm{orb}=-GM_1M_2(1-e)/(2r_p)$ and 
orbital angular momentum  $|\mathbf{L}_\mathrm{orb}|=M_1M_2\sqrt{G r_p(1+e)/(M_1+M_2)}$ throughout the simulation,
{where $M_1=M_{*,1}+M_{d,1}$ and $M_2=M_{*,2}+M_{d,2}$. This 
definition of orbital angular momentum
is equivalent to separating the total angular momentum of the system 
 $\mathbf{L}_\mathrm{tot}$ obtained by direct summation (disc cells and stars)
 into two components:
 the inner angular momentum of each disc and its host star $\mathbf{L}_\mathrm{disc}$
(measured with respect to the centre of mass of each star-plus-disc system, which is roughly the location of the star) 
and $\mathbf{L}_\mathrm{orb}$.}
Figure~\ref{fig:energy_results} shows  $E_\mathrm{orb}$ normalized
by $GM_1M_2/r_p$ (left column) and  $|\mathbf{L}_\mathrm{orb}|$ normalized by
its value at $t=0$. These figures share the same axes range to highlight the dramatic
differences in energy and angular momentum change in the orbits. While \texttt{`PARA1'} and \texttt{`PARA2'}
show substantial evolution, sets \texttt{`PARA3'} through \texttt{`PARA5'} change
their orbital properties by very small amounts. {The \texttt{`PARA4' and} \texttt{`PARA5'} runs
also show good conservation of total angular momentum since tidal effects are small, and therefore
triggered accretion is minimal (recall that some angular momentum can be lost owing
to the accretion of rotating gas onto the stars). The change in $L_{\mathrm{tot},z}$ over $\sim3000$ years in
the  \texttt{`PARA5'} runs ranges from $-1\%$  to $+0.5\%$, which is as constant as observed in isolated
discs.}

The shaded region in 
Figure~\ref{fig:energy_results}  defines the ``interaction period" outside of
which the tidal forces are expected to have very little effect. This window is defined as proportional to the pericenter timescale
$t_\mathrm{peri}\equiv r_p/v_p=r_p/\sqrt{\mu/r_p}$, where we use $\mu = G\times 1 M_\mathrm{tot}$ with $M_\mathrm{tot}=1\,M_\odot$.
Empirically, we find that a window of half-width equal to $6\times t_\mathrm{peri}$ encloses
most of the energy and angular momentum change centered around pericenter passage $\tau_0$. In practice,
the interaction window has a width of $\sim40$~yrs for \texttt{`PARA1'} (i.e., consisting of only a handful
of snapshots) and of $\sim4600$~yrs for \texttt{`PARA5'}, which covers nearly the full integration.
Most importantly, the asymmetry of the total integration time with respect
to pericenter time $\tau_0$ implies that for very long interaction periods with half-lengths $\gtrsim\tau_0$, the tidal interaction
preceding proper pericenter is not entirely captured by the simulation. This is the case in
\texttt{`PARA4'} and \texttt{`PARA5'}, for which the tidal interaction is expected to begin
at wider separations than the ones included in our initial conditions (see discussion on the fractional
error in orbital angular momentum in Section~\ref{sec:orbits}). It is possible that some of the 
differences between \texttt{`PARA4'} and \texttt{`PARA5'} as compared  to the other orbital tests (for example, that
there is no steep jump in eccentricity right before pericenter in Figure~\ref{fig:energy_results})
could be due to an extremely wide tidal interaction window and that
\texttt{`PARA4'} and \texttt{`PARA5'} are simply ``incomplete", that is,
their integration should have begun at greater separations in order to cover the asymptotic interaction.

\subsection{Comparison to linear theory}
The tidal interaction between a star+disc system and another stellar flyby was studied in detail
by \citet{ost94}. Although that work focused on a simpler system containing only one disc, a comparison
should be meaningful for our wide-separation simulations. 

\begin{figure}
\centering
\includegraphics[width=0.48\textwidth]{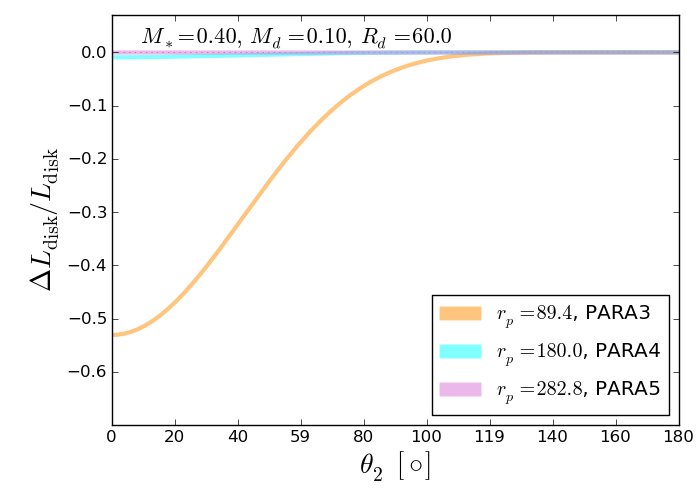}
\caption{Change in a disc's internal angular momentum according to Equation~(\ref{eq:ostriker_formula})
\citep{ost94} evaluating the physical and orbital parameters of simulation sets  \texttt{`PARA3'}, \texttt{`PARA4'} and \texttt{`PARA5'}.
\label{fig:ostriker_formula}}
\end{figure}

Assuming one disc changes orientation (in our simulations, the first five runs of each subset
only differ in the value of $\theta_2$) the angular momentum loss suffered by the victim disc is, in the linear
regime (Eq. 3.1 in \citealp{ost94}),
\begin{equation}\label{eq:ostriker_formula}
\begin{split}
\Delta L_\mathrm{disc,2}&=-\mathcal{C}\left[\cos\frac{\theta_2}{2}\sin\frac{\theta_2}{2}\right]^4\\
&~~~~~\left[\frac{2}{\Omega(R_d)}\right]\left(\frac{3}{2}\right)^2\left[\cos\frac{\theta_2}{2}\Big/\sin\frac{\theta_2}{2}\right]^4
\end{split}
\end{equation}
where the normalization factor
\begin{equation}
\nonumber
\begin{split}
\mathcal{C} = &2^3\pi^2{GM_{*,2} R_d \Sigma(R_d)}\left(\frac{M_{*,1}+M_{d,1}}{M_\mathrm{tot}}\right)\\
&\times\exp\left[-\frac{2^{5/2}}{3}\left(\frac{M_{*,2}}{M_\mathrm{tot}}\right)^{1/2}\left(\frac{r_p}{R_d}\right)^{3/2}\right]
\end{split}
\end{equation}
is a rapidly decreasing function of the ratio $r_p/R_d$. Equation~(\ref{eq:ostriker_formula})
is only valid when $r_p>R_d$, which is satisfied by our simulation sets \texttt{`PARA3 - PARA5'},
with \texttt{`PARA3'} being a marginal case, since the two discs overlap near pericenter. 
Figure~\ref{fig:ostriker_formula} shows the fractional change in disc angular momentum 
(Equation~\ref{eq:ostriker_formula} normalized by $L_\mathrm{disc} $)
evaluated for our disc model with $M_*=0.45M_\odot$, $M_d=0.05M_\odot$ and $R_d=60.0$~AU. 
For  \texttt{`PARA4'} and \texttt{`PARA5'}, the victim disc experiences changes in angular
momentum that are at the level of $1\%$ and below. Since $L_\mathrm{disc} /|\mathbf{L}_\mathrm{orb}|\sim0.03$
and $0.02$ for  \texttt{`PARA4'} and \texttt{`PARA5'} respectively, the angular momentum 
exchange between the disc and the orbit is of the order of $10^{-4}$ times smaller than 
$|\mathbf{L}_\mathrm{orb}|$ which is more than an order of magnitude smaller than the change we observe
in Figure~\ref{fig:energy_results}. This suggests that the orbital evolution of  \texttt{`PARA4'} and 
\texttt{`PARA5'} is either dominated by accretion, noise or perhaps amplified by the presence of 
a second disc. On the other hand, \texttt{`PARA3'} shows a loss in inner angular momentum
of $40\%$ for nearly prograde encounters and a negligible gain during retrograde encounters. 
Since in \texttt{`PARA3'} $L_\mathrm{disc} /|\mathbf{L}_\mathrm{orb}|\sim0.04$, the victim disc is expected to 
loose angular momentum to the binary orbit by an amount of the order of $2\%$ of $|\mathbf{L}_\mathrm{orb}|$. Although
this is the correct order of magnitude for the change in $|\mathbf{L}_\mathrm{orb}|$ observed for \texttt{`PARA3'} (Figure~\ref{fig:energy_results}), this
change in orbital angular momentum comes in the form of a {\it loss} and not a gain. Again, this suggests other mechanisms are at play 
in addition to tidal forces, and that the simulation set \texttt{`PARA3'} is outside the regime represented by the work of \citet{ost94}. Interestingly,
only \texttt{`PARA1'} and \texttt{`PARA2'} show statistically significant gains in angular momentum in some of their examples. However, since
at such close encounters the discs collide violently, the departure from the linear regime is too great for sensible interpretation.

\begin{figure*}
\vspace{-0.12in}
\centering
\includegraphics[width=0.84\textwidth]{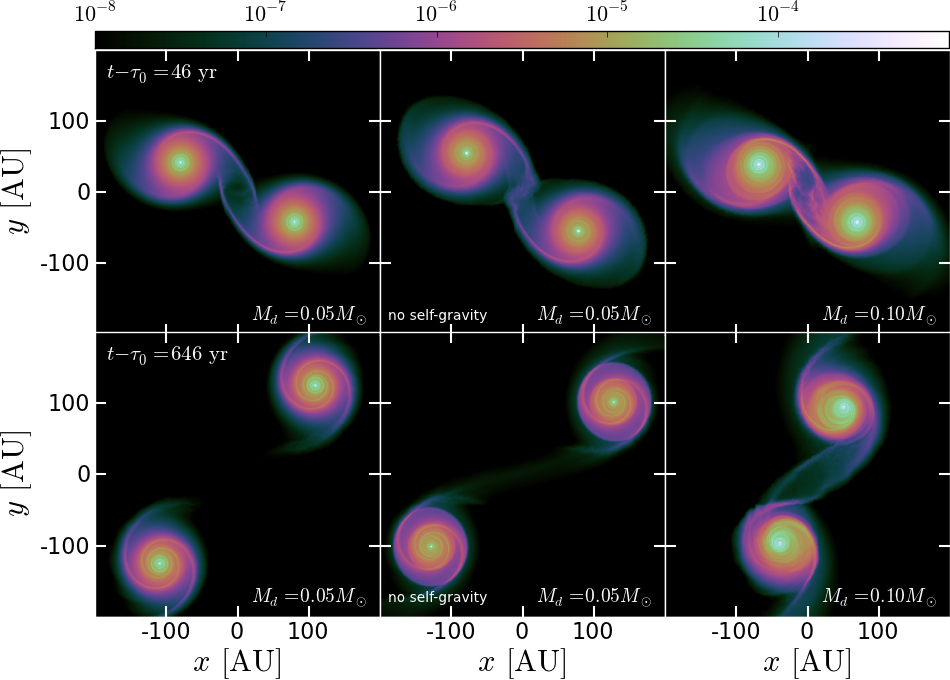}
\vspace{-0.1in}
\caption{Encounter with configuration \texttt{`PARA4-1'} ($r_p=180.0$, $\theta_1=0$, $\theta_2=0$) at two different times for three different mass scalings.
Left panel: encounter with mass ratio of $M_d/M_*=0.05/0.45$ (one of the main simulations listed in Table~\ref{tab:simulations}).
Middle panel: simulation \texttt{`PARA4-1'} but with self-gravity turned off, i.e., effectively a mass ratio of $M_d/M_*=0$.
Right panel: encounter  $M_d/M_*=0.1/0.4$.
\label{fig:encounter_highmass1}}
\end{figure*}

According to Equation~(\ref{eq:ostriker_formula}), doubling the mass of the disc while 
keeping $M_*+M_d$ constant produces an increase by only $\sim20\%$ in $\Delta L_\mathrm{disc} $,
thus requiring very massive discs in order for tidal effects to have a significant impact. We have tested the effect of doubling the disc mass
while keeping the sum $M_*+M_d=0.5M_\odot$ constant and run the set of orbits  \texttt{`PARA4'} at these higher masses. 
Figure~\ref{fig:encounter_highmass1} shows a comparison of the run  \texttt{`PARA4-1'}  with masses 
$M_*=0.45M_\odot$ and $M_d=0.05M_\odot$, and 
$M_*=0.4M_\odot$ and  $M_d=0.1M_\odot$. Although the more massive discs show hints for
richer inner structure, presumably triggered by the lower value of the Toomre $Q$ parameter, 
{(in this case, $Q_\mathrm{T,min}\approx2$)}, 
the orbital evolution seems to be mildly altered by the change in mass.
In addition, a simulation with self-gravity artificially turned off is shown for comparison purposes.

Increasing the disc mass dramatically changes the outcome for non-aligned discs. Figure~\ref{fig:encounter_highmass2} 
shows the same set-up as Figure~\ref{fig:encounter_highmass1} but now with the orientations corresponding to \texttt{`PARA4-4'}. 
One can see that the stars are rapidly captured after pericenter passage, in striking contrast to the case \texttt{`PARA4-1'}.
In this example, the pericenter separation is larger than twice the disc size (in this case, $r_p=180$~AU) and therefore the 
discs do not collide directly. Because of this, the change in the stars' orbital evolution is not due to radiative losses
as in the \texttt{`PARA1'} cases, but instead due to conservative tidal forces (at least during the early phases of the interaction). 
As one can expect, the non-axisymmetric tidal features are now more massive in proportion to the discs' increased mass,
and therefore can torque the stellar trajectories more effectively. Furthermore, by inclining the discs, the warping modes
triggered in the discs make the tidal torque they exert on the stars even stronger \citep[this can already be seen in
the linear regime calculations of][]{ost94}. As opposed to the limit of  small disc mass, in which
$\mathbf{L}_\mathrm{tot}\approx\mathbf{L}_\mathrm{orb}\sim$constant, massive discs with different inclinations can
contribute a significant amount to $\mathbf{L}_\mathrm{orb}$, making it no longer directed along the $z$-direction.
{For these massive discs, both the disc orientation and the binary orbital plane can be torqued as a result of the encounter}. This mutual torquing is 
strengthened with larger disc mass and with higher mutual inclination, favoring the disruption of the initially
parabolic orbits with greater efficiency, even within the short timescales ($\sim1$ orbital period) of the simulation.

This work opens interesting possibilities for the outcome of disc-disc interactions well into the non-linear regime. Future work
should explore the role of energetics in more detail, studying the interplay between mechanical and thermal energy in the discs,
and how realistic cooling prescriptions within the violent compression shocks can affect the results found here.

\section{Discussion: Are disc encounters likely in star-forming regions?}~\label{sec:discussion}
In this work, we have shown that realistic
circumstellar discs can affect the orbital evolution of their host stars {\it provided} the encounters
have pericenter separations close to or smaller than the disc sizes. The orbital evolution changes dramatically only
for pericenter passages comparable to the disc sizes. Wider encounters may still be important for mis-aligned, massive discs.

Fly-by encounters between young stars at impact parameters of $\sim100$~AU (all encounters
in our numerical experiments except for the most extreme case \texttt{`PARA1'} for which the
impact parameter is set to 50~AU) are likely in dense stellar clusters \citep[see][]{thi05} 
provided cluster dispersion does not happen on timescales much shorter than the mean encounter
timescale. More generally the cluster lifetime is an important variable since open clusters in the Galactic plane evaporate
on timescales of one to a few crossing timescales, or $\sim1$ to 3 Myr \citep{reg11,jef11} with a small fraction reaching up to 10 Myr. 

\begin{figure*}
\centering
\includegraphics[width=0.84\textwidth]{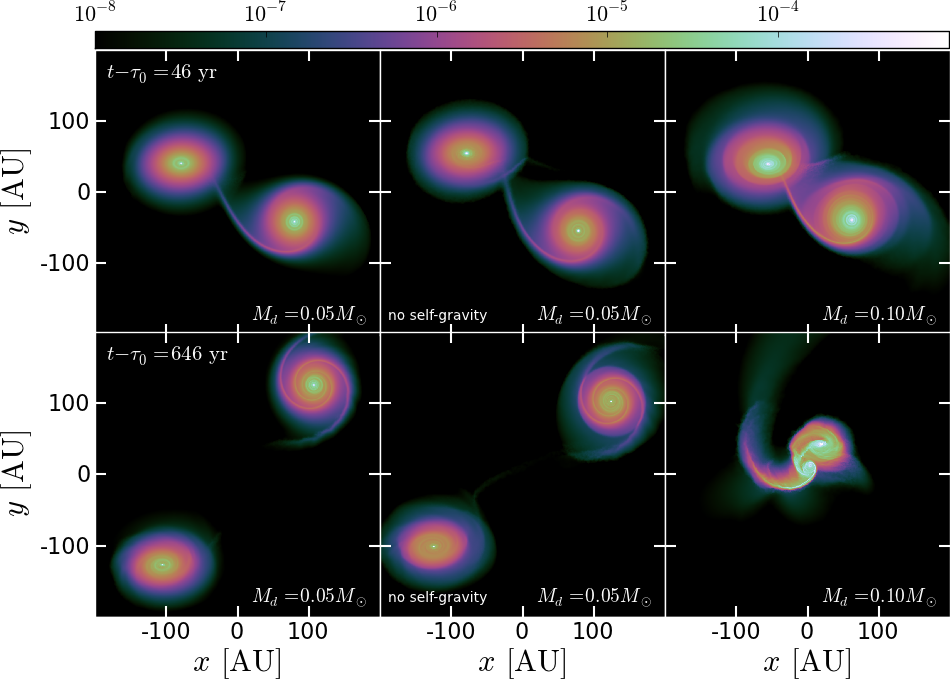}
\vspace{-0.1in}
\caption{Same as Figure~\ref{fig:encounter_highmass1} but with disc orientations as those of
simulation \texttt{`PARA4-4'} ($r_p=180.0$, $\theta_1=0$, $\theta_2=135$).
\label{fig:encounter_highmass2}}
\end{figure*}

The two-body encounter rate follows a Poisson process if events are not
correlated in time. In this case, the probability for one star to suffer
{\it at least} one encounter occurring in any interval of length $\Delta t$
is $P_{\Delta t}=1-\exp\left({-\Gamma\Delta t}\right)$,
where the encounter rate $\Gamma$ is $\Gamma(b,v)=\pi n b^2 v$
for a given impact parameter and encounter velocity $v\sim$
the cluster dispersion velocity $\sigma_v$. For a spherical cluster, 
the number density, velocity dispersion and crossing time are respectively:
 $n={3\pi^{1/2}\Sigma_c^{3/2}}/(´{4\bar{m}M_c^{1/2}})$~,
$\sigma_v\approx G^{1/2}(\pi\Sigma_cM_c)^{1/4}$~ and
$\tau_\mathrm{cross}={M_c^{1/4}}{G^{-1/2}(\pi\Sigma_c)^{-3/4}}$~
\citep[e.g.,][]{duk12}, where $\bar{m}$ is the median stellar mass.

For a  period of time corresponding to
the lifetime of circumstellar discs (say $\tau_\mathrm{disc} \lesssim10$~Myr), the 
probability $P_{\Delta t=\tau_\mathrm{disc} }$ can be computed as a function of impact parameter $b$
for a given cluster mass $M_c$ and surface density $\Sigma_c$ \citep[see][for a similar analysis]{thi05}.
For clusters with surface densities of $\Sigma_c=1.0$~g~cm$^{-2}=4.7\times10^3$~$M_\odot$~pc$^{-2}$
and masses of $M_c$ between $10^{1.5}$ and  $10^{2.5}$~$M_\odot$ the encounter probability per
star can be up to $0.6$ for $b<100$~AU and $\approx$1 at $b<400$~AU after 10~Myr. For 
$\Sigma_c=0.1$~g~cm$^{-2}$ the encounter probabilities are still high, $0.01$ to $0.02$ at 100~AU and
0.1 to 0.2 at 400~AU.

Young clusters disperse after gas expulsion, and the number density
of objects decreases rapidly after a few crossing times \citep[e.g.][]{elm00,har01,lad03}.
Since $\tau_\mathrm{cross}$ will almost invariably be shorter than $\tau_\mathrm{disc} $,
the realistic probability of encounter has an upper bound of $P_{\Delta_t=\tau_\mathrm{cross}}$.
In this case, the likelihood of an encounter
depends only on $\Sigma_c$ (this was pointed out by \citealp{duk12}, but see \citealp{cra13} 
for a discussion on the validity of this result). If the waiting time is only one crossing time,
the chance of an encounter as a function of $b$  becomes two orders
of magnitude smaller than the previous estimate. For the impact parameters
explored in our numerical simulations (50, 100, 200, 300 and 400 AU before gravitational focusing)
the probability of a star undergoing such encounters is roughly $10^{-4}$, $5\times10^{-3}$, 
$10^{-3}$, $5\times10^{-2}$ and $10^{-2}$, respectively, for the values of $\Sigma_c$ considered.
For a median stellar mass of $\bar{m}\approx0.6\,M_\odot$, the range of cluster masses
considered here imply a few tens to a few hundred members. Thus, 
for the most optimistic cluster properties, we expect a global number
of encounters per cluster of order unity.  This can be expressed by the number of encounters per star
expected within a crossing time. A simple estimate gives
\begin{equation}\label{eq:expected_encounters}
N_\mathrm{enc,exp}\approx\langle\Gamma(b,v)\rangle\times t_\mathrm{cross}=\frac{3}{4}\pi b^2 \frac{\Sigma_c}{\bar{m}}~,
\end{equation}
which is  a small number for the cluster parameters explored above.

Thus encounters which dramatically reshape binary orbits are rare, except perhaps for the widest configurations we 
have considered with
(the \texttt{`PARA5'} case).
In this case, flybys could still play a role in disc truncation, but owing to tidal forces
associated with the central stars rather than ram pressure stripping as is observed in
the head-on disc collisions. For the  \texttt{`PARA4'} and \texttt{`PARA5'} cases 
the effect of the finite disc mass on the stellar orbits is negligible, unless the disc
masses are increased from $M_d=0.11M_*$ (fiducial case) to, e.g., $M_d=0.25M_*$.
Including large disc masses is the only way in which discs might play a dynamical role in the evolution
of the cluster at small scales due to stellar fly-bys. Since discs in fly-bys should have
uncorrelated angular momentum vectors, such encounters would be 
most similar to the third panel in Figure ~\ref{fig:encounter_highmass2}.

Whether such massive discs are realistic is uncertain. Class 0 sources are thought to have up to $85\%$ \citep{loo03} of the mass of the system
in the envelope, but it is unclear how much of that mass is in the form of a disc, since both the disc and signatures of Keplerian
rotation are difficult to detect due to instrumental and optical depth limitations.
\citet{jor09} estimated through indirect methods that the
submillimeter continuum emission from protostars was consistent with discs having masses
as high as $0.5\,M_\odot$, however direct evidence of Keplerian rotation was not
detected in their Class 0 sample. In contrast, independent modeling \citep[e.g.][]{chi08} suggests
that disc masses in Class 0 objects could be comparable to T Tauri star discs.
Of those Class 0 sources for which there is an unambiguous detection of Keplerian rotation,
thus far only  L1527 has a (model dependent) estimate of a disc mass, which is $\lesssim1\%$
of the protostar mass \citep{chi12,tob13}. {But, even in this case, evidence may
not be conclusive, as optical-depth effects may underestimate the disc mass, which could be
as high as half the host star mass as argued by \citet{for13}. Such a massive disc would enable
angular momentum transport via gravitational instabilities, which in turn would be consistent with
the inferred accretion rate of this object.} In summary, although disc-to-star mass ratios of $~0.25$
provide enough torques onto the stellar orbits to produce quick orbital decay, {Observational
evidence is not yet conclusive as to whether such massive Class 0 discs exist. Future Atacama
Large Millimeter Array data will help resolve this question (Segura-Cox et al, in prep).}

If very high disc masses are not confirmed, enhancements
in the encounter rates owing to realistic  cluster substructure might lead to a greater encounter frequency, too.
As these fractal clusters undergo relaxation and dispersal, \citet{cra13} found the number of encounters  
can be one or two orders of magnitude higher than the relaxed-cluster estimate
of Equation~(\ref{eq:expected_encounters}). In fact, in their numerical
experiments \citet{cra13} find that $\gtrsim50\%$ of the close encounters 
(defined by those authors as all those within $b=1000$~AU) take place within 
a crossing time, before the rates go down to values 
consistent with Equation~(\ref{eq:expected_encounters}).
For some of their densest cluster models, these authors find that the 
number of encounters per star can be in the hundreds. Assuming that the probability 
distribution of impact parameters is still described by
$p(b)db\propto b\,db$, encounters at  $b\sim500$~AU will be four times less likely
than at 1000~AU, and those at  $b\sim100$~AU will be 100 times less likely, still
allowing for a few encounters per star to take place before the cluster disperses. 
These considerations expand the parameter space available for plausible disc encounters
to include most of the impact parameters explored in our numerical experiments.

Although the orbital decay/capture of the stars is still a sensitive function of disc/envelope mass,
the increased encounter rates given by \citet{cra13} (which are independent of disc mass)
imply that flyby disc encounters could provide a means for truncation early on in the disc
lifetime. Extreme environments like Orion nebular cluster may already provide the evidence
of this process having taken place, as suggested by \citet{jua12}.

\section{Summary}
We have carried out simulations of binary disc encounters using the moving-mesh
code {\footnotesize AREPO}. We have focused our analysis
on the effects of tidal forces on a stellar pair in a
parabolic orbit, exploring the outcome of disc-disc interactions well into the non-linear regime.

We have found that the orbital energy and orbital angular momentum change 
per passage depend sensitively on the separation at pericenter. This result 
is qualitatively consistent with tidal torque calculations of star-disc interactions
for parabolic and hyperbolic orbits \citep{ost94}. One surprising result, however,
is the outcome of ``runaway orbital decay" in those simulations with small pericenter 
separations ($r_p\sim R_\mathrm{disc} $), which show orbital capture with subsequent energy 
losses at each pericenter encounter, hastening disc dispersal and eventually 
forming close binaries with diffuse circumbinary discs. This process is particularly
fast for retrograde discs. We expect this result to be
very sensitive to the gas equation of state, and thus future work should explore the
role of cooling prescriptions or realistic radiative transfer on the dynamics of discs
affected by violent compression shocks.

The applicability of our numerical results to real astrophysical systems
remains to be assessed by upcoming high resolution observations of 
circumstellar discs. In principle, the configurations 
explored here may be too extreme to be common. Indeed, in most nearby 
Galactic star forming regions, the effective surface density of protostellar clusters 
is too low to produce more than $\sim10^{-3}$ encounters {\it per star} before
the clusters dissipates  (a timescale typically shorter than the lifetime of discs). 
However, there is some evidence of strongly interacting circumstellar discs
with morphologies that could in principle be explained by stellar flybys
\citep[see][]{cab06,sal14}. In the near future, ALMA continuum and line
data will provide a large sample of discs in binaries which will help to address
the statistical relevance of those existing candidates and measure the frequency of flyby disc
pairs. If flybys are found to be more frequent than expected from
the simple arguments above, cluster substructure is a plausible
explanation for enhanced rates  \citep[][]{cra13}.There is clear evidence that young 
clusters are not uniform, spherical, virialized systems,
but instead have significant fractal-like substructure 
\citep[in both gas and stars; see e.g.,][]{elm00,lad03,car04,gut11}.

Similarly, new data on circumbinary gas discs from
ongoing and future surveys will test different formation mechanisms
for such systems. The occurrence rate
of circumbinary discs will help provide answers to the question 
of whether these form in an analogous manner to discs around single stars,
or if their frequency is modulated by other ``non-primordial" 
formation mechanisms such as the ones produced by the simulations in this work.
 
Our experiments of isolated disc encounters allow for a detailed analysis
of disc evolution in densely packed star forming regions, where disc
truncation can take place, in contrast to global cluster simulations, for
which an accurate description of the disc dynamics may be too costly or impossible 
to implement \citep[e.g.][]{bat12,ros14}. The exploration of the role of cluster kinematics 
on disc evolution via isolated flybys
may help provide a link between the environmental properties of the parent cluster
and the planet forming capabilities of a young star. Under this
scenario, the planet forming efficiency will not only depend
on the mass, metallicity and dust content of a star and its surrounding disc, but
also on the location within the birth cluster and on the overall properties of the
latter.

\section*{acknowledgements}
{The research presented here was largely carried out as part of DJM's PhD
  thesis at Harvard University.  The simulations in this paper were
  run on the Odyssey cluster supported by the FAS Science Division
  Research Computing Group at Harvard University. DJM would like to thank Dimitar
  Sasselov, Matthew Holman, Ruth Murray-Clay and James Stone for
  insightful feedback and support throughout the development of this
  work. DJM acknowledges partial support from the Fulbright-CONICYT graduate fellowship
  program.}
  


\appendix

\section{Isolated Thin disc Models}\label{app:models}
%
\subsection{Model characteristics and initial conditions}
To obtain stable initial disc conditions we enforce radial centrifugal and vertical hydrostatic equilibrium:%
\begin{align}\label{eq:radial}
-\frac{v_\phi^2}{R} = -\frac{1}{\rho}\frac{\partial p}{\partial R} - \frac{\partial \Phi_0}{\partial R} - \frac{\partial \Phi_g}{\partial R}~~,    \\
\label{eq:vertical}
0 = -\frac{1}{\rho}\frac{\partial p}{\partial z} - \frac{\partial \Phi_0}{\partial z} - \frac{\partial \Phi_g}{\partial z}~~,
\end{align}
where $\Phi_0(R,z)$ is the potential due to the central star and $\Phi_g(R,z)$ is the potential due to gas self-gravity. 
Equation~(\ref{eq:radial}) determines the azimuthal velocity field $v_\phi^2(R,z)$ while
Equation~(\ref{eq:vertical}) contains the solution for the vertical structure of the disc at all radii.

The models used here are based on the Lynden-Bell-Pringle surface density profile 
(\citealp{lyn74}; Equation~\ref{eq:surface_density}). We use 
a fixed temperature profile of the form $T(R)=T_c \left({R}/{R_c}\right)^{-l}$ with power-law 
index fixed to $l=0.5$. The disc characteristic radius is set to $R_c=20$~AU and the total mass of the system
(star plus disc) to $0.5~M_\odot$. The disc-to-star mass ratio is varied from $M_d/M_*=0.02$ up
to $0.67$, always keeping $M_*+M_d=0.5~M_\odot$. The normalization of the temperature profile
is chosen according to a specified aspect ratio at $R_c$, producing a flared disc with aspect ratio
\begin{equation}\label{eq:aspect_ratio}
h(R)\equiv \frac{c_s(R)}{v_K(R)}=\sqrt{\frac{k_BT(R)}{\mu m_\mathrm{H}}\frac{R}{GM_*}}\equiv h_c \left(\frac{R}{R_c}\right)^{-(l-1)/2}.
\end{equation}
We vary $h_c$  from $0.04$- $0.1$.  The minimum value of $h_c$ is set by resolution constraints: for smaller values a proper three-dimensional description
requires $N_\mathrm{gas}\sim10^7$.

All singular terms in Equations~(\ref{eq:surface_density})~and~(\ref{eq:aspect_ratio}) are regularized using a spline softening 
with softening parameter $s$ 
 akin to that used for the gravitational potential of the central stars and of individual gas cells \citep{her89,spr01}.
In this case, the central gravitational potential $\Phi_0$ reaches a finite value at $R=0$ (the position of the star) and recovers its
exact Keplerian value at $R=2.8s$. This means that the Keplerian angular speed $\Omega_K^2={(\partial\Phi_0/\partial R)/R}$ vanishes at $R=0$.

\vspace{-0.1in}
\subsection{Mesh generation and vertical structure}
We tessellate the domain using a {\it mass resolution} criterion,
in which the domain is discretized by a Monte-Carlo sampling of an underlying density field. 
Given a number of mesh-generating points $N_\mathrm{gas}$, a gas disc of total mass $M_d$ is
split into cells of nearly equal mass $m_\mathrm{gas}=M_d/N_\mathrm{gas}$. 
We assume a 3-D density field of the form
$\rho(R,z)=\Sigma(R)\zeta(R,z)$, where $\int_{-\infty}^{+\infty}\zeta(R,z)dz=1$.
Because the density field is not separable into $R$ and $z$ components,
we sample the radial positions $\{R_i\}$ first
and then sample the vertical positions $\left.\{z_i\}\right|_R$ for {\it fixed} $R$
\footnote{The 1-D function $\Sigma(R)$ is sampled through conventional methods
to produce a set  $\{R_i\}$ of size $N_\mathrm{gas}$.
Assuming the variability in $R$ is slower than
that in $z$, the $\{R_i\}$ are grouped into radial bins and proceed to
Monte-Carlo sample the 1-D function $\zeta(z|R)$ to obtain values for
the $z$-coordinate.}.
The azimuthal coordinate $\phi$ is drawn from a uniform distribution in $[0,2\pi)$.

The vertical profile at a given $R$, $\zeta(z|R)$ must be obtained numerically for self-gravitating discs.
We have adapted the potential
 method described in detail in \citet{wan10}  \citep[see also][]{spr05b}.)
This technique consists of iterating 
between vertical hydrostatic equilibrium equation -- under a fixed vertical potential -- 
and  the vertical Poisson equation -- for a given vertical density profile -- until convergence 
is achieved. The geometrically thin disc approximation allows us
to solve for two coupled ordinary differential equations instead of
a set of partial differential equations. We force the  separability of the potential of a very flattened system
 into a mid-plane component and a local vertical component in
 the form $\Phi_g(R,z)=\Phi_g(R,0)+\Phi_{g,z}(R,z)$  \citep{bin08},
such that a local, 1-D Poisson equation can be written in the variable $z$ for fixed $R$.
We use a locally isothermal equation of state, in which
 the sound speed only depends on the radial coordinate $R$, such that
 $p=c_s^2(R)\rho$. This approximation, combined with Equation~(\ref{eq:vertical}) 
and  provided we know the mid-plane density $\rho(R,0)\equiv\rho_0(R)$ allows
us to solve for the vertical density profile:
\begin{equation}\label{eq:vertical_profile}
\rho(R,z)=\rho_0(R)\;\exp\left(-\frac{\Phi_{0,z}+\Phi_{g,z}}{c_s^2(R)}\right)~~.
\end{equation}
where $\Phi_{0,z}\equiv\Phi_{0}(R,z)-\Phi_{0}(0,z)$ is the $z$-dependent part of
the Keplerian potential $\Phi_{0}$ of the central star

Starting with an initial guess for the midplane density $\rho^{(0)}_0(R)$ and
for disc self-potential $\Phi_{g,z}^{(0)}=0$, Equation~(\ref{eq:vertical_profile})
can be used to solve to obtain a full density profile from which a new potential can
be derived, which in turns allows for a new estimate of the midplane density. 
The iteration steps are (for fixed $R$):
\begin{subequations}\label{eq:vertical_iteration}
\begin{align}
\mathrm{(I)}&~~~~\rho_0^{(k+1)}= \cfrac{\Sigma(R)}{\int\limits_{-\infty}^{\infty}\cfrac{}{}\rho^{(k)}(z)\,dz}\\
\begin{split}
\mathrm{(II)}&~~~~\text{solve numerically for}~~\Phi_{g,z}^{(k+1)}\\
&~~~~~\cfrac{d}{dz}\begin{pmatrix}\Phi_{g,z}^{(k+1)\sp{\prime}}\\ \\\Phi_{g,z}^{(k+1)}\end{pmatrix}
=\begin{pmatrix} 4\pi G\rho^{(k)}(z)  \\ \\ \Phi_{g,z}^{(k+1)\sp{\prime}}\end{pmatrix}
\end{split}~~\\
\mathrm{(III)}&~~~~\rho^{(k+1)}(z)=\rho_0^{(k+1)}\;\exp\left(-\frac{\Phi_{0,z}+\Phi_{g,z}^{(k+1)}}{c_s^2}\right)~~.
\end{align}
\end{subequations}
For low mass circumstellar discs ($M_d\lesssim0.05M_*$) convergence can be achieved
within at most 5-6 iterations using as the initial guess a Gaussian-profile disc. 
If discs are massive ($M_\gtrsim0.1M_*$),
convergence is achieved most quickly using the self-gravitating
slab \citep{led51} as the initial guess (even though it neglects the presence of the star's Keplerian potential).
For discs masses of $M_d\sim0.2M_*$, the latter approach needs at most $\sim9$ iterations for convergence.

For the number of mesh-generating points per disc that are used in the study,
the evaluation of the density
field is complete in the vertical direction up to  $\sim3$ scale-heights. Beyond this point, the mesh transitions into a quasi-regularly spaced mesh that 
connects to the coarse background cells that fill the computational domain (see \citealp{spr10a} for
a detailed description).

\begin{figure}
\includegraphics[width=0.42\textwidth]{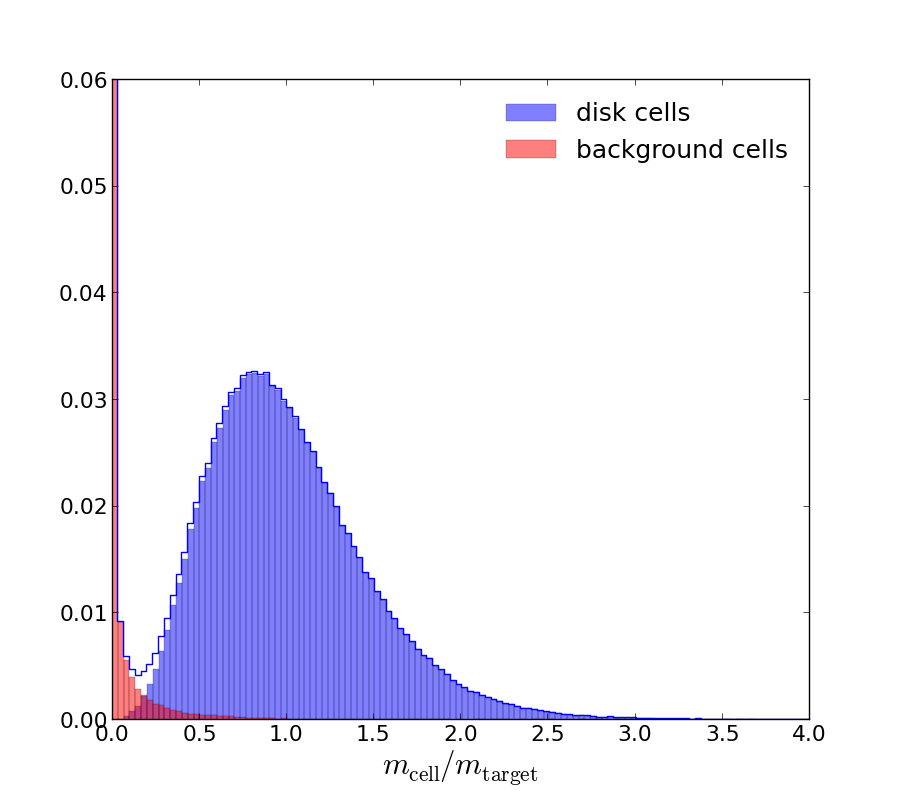}
\vspace{-0.1in}
\includegraphics[width=0.42\textwidth]{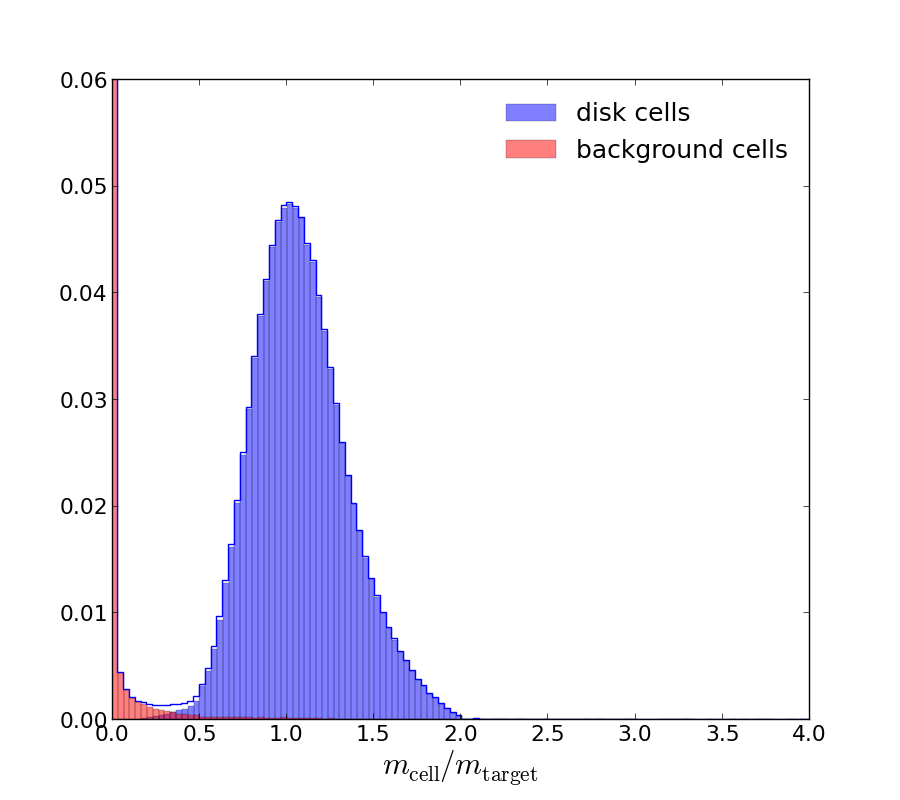}
\caption{Mass distribution of Voronoi cells resulting from the tessellation
of Monte-Carlo sampled points (top panel) and the distribution
resulting from refinement/derefinement of cells above/below the
mass limit $0.5$/$2.0$ times the reference mass $m_\mathrm{target}$ (bottom panel).
\vspace{-0.2in}
\label{fig:mass_distribution} }
\end{figure}

To minimize the Poisson noise of the sampling, we take advantage of the Eulerian nature  of {\footnotesize AREPO}
and impose the exact value of the density field at the mesh-generating points, instead of forcing 
an homogeneous cell mass $m_\mathrm{gas}$ throughout the disc.
This additional step -- only possible with {\footnotesize AREPO}\ -- produces very smooth density
fields, at the expense of a variation in cell mass, although this can be mitigated. 
 Figure~\ref{fig:mass_distribution} shows the distribution of cell masses as generated by the
initial condition algorithm (top panel) and how the spread in cell masses can be corrected (bottom panel) by a series of refinement
and derefinement steps \citep{spr10a}. Figure~\ref{fig:mass_distribution} also shows the distribution
of ``background cells", which are very low mass cells or large volume
that fill in the computational box. Ideally, these two cell distributions
should smoothly transition into one another (see Figure~\ref{fig:mesh_slice}).

We confirm that the prescribed surface density $\Sigma(R)$ is recovered
from the 3D models by integrating the density field along the vertical
direction.

\subsection{Velocity structure}
The velocity field of a stationary, axisymmetric system is
given by the solution for $v_\phi$ in Equation~(\ref{eq:radial}).
The velocity is not exactly Keplerian due to 
the negative contribution of the pressure gradient and disc self-gravity.
In addition, three-dimensional discs will have a small vertical gradient
in orbital speed which causes the upper layers to rotate more slowly than
the mid-plane. Qualitatively,
\begin{equation}
\begin{split}
{v_\phi^2}=&\left(\text{Keplerian term}\right)^2+\left(\text{self-gravity term}\right)^2\\
&+\left(\text{pressure term}\right)^2 + \left(\text{vertical layering term}\right)^2~~.
\end{split}
\end{equation}
For discs with masses greater than $0.02\,M_*$, the self-gravity term is
comparable to, or greater than the pressure term, although it only starts to cause
a significant deviation from Keplerian rotation for $M_d/M_*\gtrsim0.3$.

The velocity field in three dimensions,  for locally isothermal discs, 
is \citep[see][]{wan10}
\begin{equation}\label{eq:rotation2}
\frac{v_\phi^2}{R} = \frac{v_c^2}{R} 
-\left.\frac{1}{\rho}\frac{\partial p}{\partial R}\right|_{z=0} 
 -\frac{\partial c_s^2}{\partial R}\ln\left(\frac{\rho(R,z)}{\rho_0(R)}\right)~~,
\end{equation}
where $v_c^2=v_K^2(z=0)+v_{c,d}^2$ is the circular speed due to gravity (from both the central star
and the disc). Thus the vertically-layered rotation curve can be obtained by first calculating the two-dimensional rotation curve corresponding to a highly-flattened disc \citep{bin08} and then adding a correction due
to the vertical structure \citep{wan10}. 

The component of the circular speed from self-gravity has many possible functional forms \citep[see ][]{bin08}.
For numerical computations, we have found that a formula due to \citet{mes63} is particularly
useful:
\begin{equation}\label{eq:circ_vel1} 
\begin{split}
v_{c,d}^2(R) = G\frac{M_d(<R)}{R}+\\
2G\sum_{k=1}^\infty\alpha_k&\left[\frac{(2k+1)}{R^{2k+1}}
\int_0^R\,dR'\,\Sigma(R'){R'^{2k+1}}\right.\\
&-2kR^{2k}\left.\int_R^\infty\,dR'\,\frac{\Sigma(R')}{R'^{2k}}\right]
\end{split}
\end{equation}
Although the summation in Equation~(\ref{eq:circ_vel1})  does not converge rapidly, it is
a convenient representation of $v_{c,d}^2$ for numerical computation, since it involves only well-behaved integrals and simple
sums. If a tolerance parameter magnitude of $10^{-6}$ is introduced relative to the zeroth order term ($GM_d(<R)/R$),
only the first $\sim10$ terms in the sum are necessary. In pathological cases where more terms are required, the sum
is extended to $\sim20$. This typically happens for small $R$, where often times the inaccuracy in $v_{c,d}^2$ is 
negligible relative to the dominant Keplerian term $v_K^2$.

For a density profile index of $p=1$,the self-gravity contribution to the disc's rotation
 curve can be calculated exactly (see Appendix~\ref{app:vel_profile}).
Although the analytic solution to $v_{c,d}^2(R)$ is too complicated to be useful in
practice (it involves Meijer functions), it can be used for debugging purposes and to 
compare to the numerical output of  the truncated series in Equation~(\ref{eq:rotation2}).
This solution can be written
\begin{equation}\label{eq:selfgrav_vel}
v_{c,d}^2(R) = \frac{GM_d}{R_c}f\left(\tfrac{R}{R_c}\right)~~,
\end{equation}
where $f(x)$ is a dimensionless function\footnote{
The functional form of this dimensionless function is
\begin{equation}
\begin{split}
f(x) = \frac{x}{2\sqrt{\pi}}G^{21}_{13}
\left(\frac{x^2}{4}\left| \begin{array}{c} 0 \\ -\frac{1}{2},\;\; \frac{1}{2},\;\; -\frac{1}{2}\end{array}\right.\right)~~.
\end{split}
\end{equation}
} 
of order unity which involve Meijer-G functions (Appendix~\ref{app:vel_profile}) and that equals 1
at $R=0$ and approaches $R_c/R$ as $R\rightarrow\infty$ (i.e. the disc potential
approaches a Keplerian form at large distances). 

\begin{figure}
\includegraphics[width=0.45\textwidth]{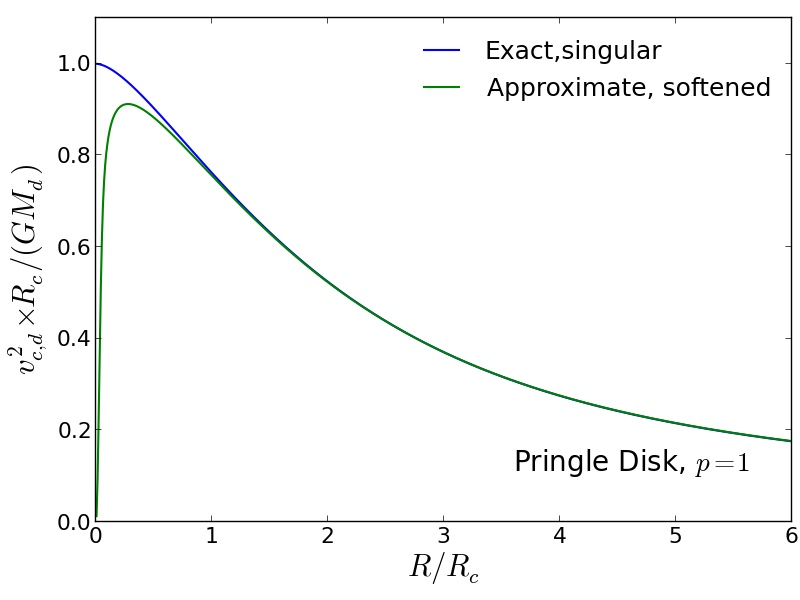}
\caption{Circular speed profile of a Lynden-Bell-Pringle disc owing to its
own self-gravity. The blue line depicts the exact computation of $v_{c,d}^2$
(Equation~\ref{eq:selfgrav_vel}) for the singular profile $\Sigma(R)$
of Equation~(\ref{eq:surface_density}) for a power-law index $p=1$ normalized
by $GM_d/R_c$. The green line shows the self-gravity contribution to
$v_\phi$ we actually use in our models (calculated through a truncated version of
Equation~(\ref{eq:circ_vel1}) by numerical integration using a version of
$\Sigma(R)$ softened at the origin.
\label{fig:exact_vel_profile} }
\vspace{-0.15in}
\end{figure}

However, since $\Sigma(R)$ in Equation~(\ref{eq:surface_density}) diverges at the origin, the surface
density there is infinite. The derivative of the gravitational field due to this
mass distribution \citep[][Eq.~2.188]{bin08} takes a finite value of $GM_d/R_c$ at the origin,
which is an unphysical value for the circular speed. Any continuous distribution of
matter should have a vanishing circular speed at the origin, and this can be obtained by
smoothing out the singular term in $\Sigma(R)$ as explained above. This produces a circular
speed profile that is zero at the origin, but rapidly converges to the exact profile
far from $R=0$. This is shown in Figure~\ref{fig:exact_vel_profile}, where the numerically
obtained form of the circular speed (Equation~(\ref{eq:circ_vel1}) changes continuously from
zero to the profile given by  Equation~(\ref{eq:selfgrav_vel}).

\section{Exact Rotation Curve for a Massive Lynden-Bell--Pringle disc with \lowercase{$p=1$}}\label{app:vel_profile}
In the special case of $p=1$, the Lynden-Bell--Pringle surface density profile takes the form
\begin{equation}\label{eq:simple_profile}
\Sigma(R)=\frac{M_d}{2\pi R_c^2}\left(\frac{R}{R_c}\right)^{-1}\exp\left[-\frac{R}{R_c}\right]~~.
\end{equation}
The circular speed of a disc with this surface density profile can be written in terms of Bessel functions \citep{bin08}
\begin{equation}\label{eq:circ_vel2}
v_{c,d}^2(R) = {2\pi G}R\int_0^\infty\,dk\,kJ_1(kR)\int_0^\infty dR' R'\Sigma(R')J_0(kR')~~.
\end{equation}
With $p=1$, the innermost integral in Equation~(\ref{eq:circ_vel2} can be
computed with the aid of the identity
\begin{displaymath}
\begin{split}
\int_0^\infty e^{-\alpha x}J_\nu(\beta x)dx
=  \frac{\beta^{-\nu}\left[\sqrt{\alpha^2+\beta^2}-\alpha\right]^{\nu}}{\sqrt{\alpha^2+\beta^2}}
\end{split}
\end{displaymath}
\citep[][$\S6.611$]{gra00}, with $\alpha=1/R_c$, $\beta=k$ and $\nu=0$. This leaves
\begin{displaymath}
\begin{split}
v_{c,d}^2(R) = {GM_d}\frac{R}{R_c}\int_0^\infty\,dk\,k \frac{J_1(kR)}{\sqrt{R_c^{-2}+k^2}}~~.
\end{split}
\end{displaymath}
which can be computed easily if we reformulate the integrand in terms of Meijer-G functions. Using
\begin{displaymath}
J_\nu(\gamma x)=G^{10}_{02}\left(\frac{\gamma^2x^2}{4}\left| \begin{array}{c}  \\ \frac{1}{2}\nu,\; \;\; -\frac{1}{2}\nu \end{array}\right.\right) 
\end{displaymath}
\citep[][$\S9.34$]{gra00}, and
\begin{displaymath}
\begin{split}
\int_0^\infty x^{\rho-1}(x+\beta)^{-\sigma}G^{mn}_{pq}
\left(\alpha x\left| \begin{array}{c} a_1,...,a_p \\ b_1,...,b_p \end{array}\right.\right) dx \\
=  \frac{\beta^{\rho-\sigma}}{\Gamma(\sigma)}G^{m+1,n+1}_{p+1,q+1}
\left(\alpha \beta \left| \begin{array}{c} 1-\rho,a_1,...,a_p \\ \sigma-\rho,b_1,...,b_p \end{array}\right.\right) 
\end{split}
\end{displaymath}
\citep[][$\S7.811$]{gra00}, we obtain
\begin{equation}\label{eq:circ_vel3}
\begin{split}
v_{c,d}^2(R) = \frac{GM_d R}{2\sqrt{\pi} R_c^2}G^{21}_{13}
\left(\frac{R^2}{4R_c^2}\left| \begin{array}{c} 0 \\ -\frac{1}{2},\;\; \frac{1}{2},\;\; -\frac{1}{2}\end{array}\right.\right)~~.
\end{split}
\end{equation}
Note that
\begin{displaymath}
v_{c,d}^2(R)\xrightarrow[R\to0]{} \frac{GM_d}{R_c}~~,
\end{displaymath}
i.e., the rotation curve has a non-zero value at the origin. This is due to the divergent surface density at $R=0$ in
Equation~(\ref{eq:simple_profile}), therefore, a softened surface density profile is needed to reproduce a physically
plausible self-gravitating rotation curve that increases from zero at the origin. This explains why the surface density
softening cannot be independent from the gravitational softening of the central star, since it must be guaranteed that
near $R=0$ the rotation curve of the disc is entirely dominated by the point mass.


 \end{document}